\documentclass[fleqn,usenatbib]{mnras}
\usepackage[T1]{fontenc}
\usepackage{ae,aecompl}
\usepackage{pdflscape}

\usepackage{graphicx}	% Including figure files
\usepackage{amsmath}	% Advanced maths commands
\usepackage{amssymb}	% Extra maths symbols
\usepackage{bm}
\usepackage{mathtools}
\usepackage{float}
\usepackage{array, makecell}
\usepackage{booktabs}
\usepackage{hyperref}
\usepackage{xspace}
\usepackage{xcolor}
\usepackage{rotating}
\usepackage{newtxtext,newtxmath}

\renewcommand{\vec}[1]{{\bm{#1}}}
\newcommand{\oper}[1]{{\bm{\mathsf{#1}}}}
\newcommand{\software}[1]{{\textsc{#1}}}
\newcommand{\noisecov}[0]{\oper{C}^{-1}}

\newcommand{\imcov}[0]{\tilde{\oper{C}}_x^{-1}}

\newcommand{\dft}[0]{\oper{D}}

\newcommand{\lensop}[0]{\oper{L}}
\newcommand{\source}[0]{\vec{s}}
\newcommand{\regul}[0]{\oper{R}_\source}

\newcommand{\msol}[0]{\oper{A}}

\newcommand{\smp}[0]{\source_\mathrm{MP}}
\newcommand{\lams}[0]{\lambda_{\source}}
\newcommand{\data}[0]{\vec{d}}
\newcommand{\model}[0]{\vec{m}}
\newcommand{\dirtyim}[0]{\data_x}
\newcommand{\noise}[0]{\vec{n}}
\newcommand{\hyp}[0]{\mathrm{H}}
\newcommand{\etalens}[0]{\vec{\eta}_\hyp}
\newcommand{\nvis}[0]{N_\mathrm{vis}}
\newcommand{\npix}[0]{N_\mathrm{pix}}
\newcommand{\nsource}[0]{N_\mathrm{src}}
\newcommand{\logdet}[0]{\log \det}

\newcommand{\logev}[0]{\log\mathcal{E}}

\newcommand{\jzsfo}[0]{MG J0751+2716\xspace}
\newcommand{\sersic}[0]{S\'{e}rsic\xspace}

\newcommand{\figscale}{0.75}
\newcommand{\cornerscale}{0.85}

\newcommand{\revisions}[1]{#1}
\title[\jzsfo lens model comparison]{A lensed radio jet at milli-arcsecond resolution I: Bayesian comparison of parametric lens models}

\author[D.M. Powell et al.]{Devon M. Powell,$^{1}$\thanks{E-mail: dmpowell@mpa-garching.mpg.de}
Simona Vegetti,$^{1}$
J. P. McKean,$^{2,3}$
\newauthor
Cristiana Spingola,$^{4}$
Hannah R. Stacey,$^{1}$
and Christopher D. Fassnacht$^{5}$
\\
$^{1}$Max Planck Institute for Astrophysics, Karl-Schwarzschild-Stra\ss{}e 1, 85748 Garching bei M\"unchen, Germany\\
$^{2}$Kapteyn Astronomical Institute, University of Groningen, PO Box 800, NL-9700 AV Groningen, The Netherlands\\
$^{3}$ASTRON, Netherlands Institute for Radio Astronomy, PO Box 2, NL-7990 AA Dwingeloo, The Netherlands\\
$^{4}$INAF $-$ Istituto di Radioastronomia, via Gobetti 101, I$-$40129, Bologna, Italy\\
$^{5}$Department of Physics and Astronomy, UC Davis, 1 Shields Ave., Davis, CA 95616, USA\\
 \\
}

\date{Accepted 2022 August 16. Received 2022 August 12; in original form 2022 July 08}

\pubyear{2019}

\begin{document}
\label{firstpage}
\pagerange{\pageref{firstpage}--\pageref{lastpage}} 
\maketitle

\begin{abstract}
We investigate the mass structure of a strong \revisions{gravitational} lens galaxy at $z=0.350$, taking advantage of the milli-arcsecond \revisions{(mas)} angular resolution of very long baseline interferometric (VLBI) observations. In the first analysis of its kind at this resolution, we jointly infer the lens model parameters and pixellated radio source surface brightness. We consider several lens models of increasing complexity, starting from an elliptical power-law density profile. We extend this model to include angular multipole structures, a separate stellar mass component, additional nearby field galaxies, and/or a generic external potential. We compare these models using their relative Bayesian log-evidence (Bayes factor). We find strong evidence for angular structure in the lens; our best model is comprised of a power-law profile plus multipole perturbations and external potential, with a Bayes factor of $+14984$ relative to the elliptical power-law model. It is noteworthy that the elliptical power-law mass distribution is a remarkably good fit on its own, with additional model complexity correcting the deflection angles only at the $\sim5$ mas level. We also consider the effects of added complexity in the lens model on time-delay cosmography and flux-ratio analyses.  We find that an overly simplistic power-law ellipsoid lens model can bias the measurement of $H_0$ by $\sim3$ per cent and mimic flux ratio anomalies of $\sim8$ per cent. Our results demonstrate the power of high-resolution VLBI observations to provide strong constraints on the inner density profiles of lens galaxies.
\end{abstract}

\begin{keywords}
gravitational  lensing:  strong --  methods: data analysis -- radio  continuum:  general -- quasars: individual: MG J0751+2716 -- galaxies: structure
\end{keywords}

%%%%%%%%%%%%%%%%% BODY OF PAPER %%%%%%%%%%%%%%%%%%

\section{Introduction} \label{sec:intro}

The density structure of the inner few \revisions{kpc} in galaxies is of fundamental interest in astrophysics, as it is shaped by a wide range of interacting physical processes. While simulations containing only cold dark matter (CDM) produce an $\rho \propto r^{-1}$ dependence in density \citep{nfw1996}, the inclusion of both baryonic processes and dark matter in models significantly complicates the picture.  

For instance, it has been shown that adiabatic contraction driven by gas accretion can produce steep inner density profiles \citep{blum1986,gnedin2011,schaller2015}. However, stellar feedback, AGN feedback, and/or subhalo accretion can alternatively beget cored profiles \citep{rd2008,pontzen2012,martizzi2013}.  Alternative models for dark matter also have an impact, with both self-interacting dark matter (SIDM), and fuzzy dark matter (FDM) exhibiting a tendency to form cores \citep{yoshida2000, vogelsberger2014b, burkert2020}.  Simulations \revisions{that} include both SIDM and baryons exhibit a complex interplay between the baryons and dark matter, and can form either cored \emph{or} cuspy profiles, depending on the age and mass accretion history of the galaxy \citep{despali2019,vargya2021}.  In addition to the complex picture of the nature and causes of their radial density profiles, galaxies also exhibit angular structure beyond perfect ellipticity \citep{bender1987,bender1988,bender1989,peng2002}.  Environment can have a strong effect in this regard. For example, the presence of ``disky'' or ``boxy'' isophotal shapes is associated with the properties of the progenitors in major mergers \citep{khochfar2005,naab2006, kormendy2009}.  Tidal interactions are also observed to play a role, with an excess of boxy or irregular galaxy shapes observed in compact groups \citep{nieto1989,zepf1993}. 

It is clearly in our interest to obtain the best possible understanding of the mass distribution in the inner $\sim 1$ \revisions{to 2 kpc} of galaxies, as this is a direct window into the processes that shape them. Strong gravitational lensing has long been regarded as an indispensable tool in this endeavor.  Its sensitivity solely to the gravitational field of the lens makes strong lensing a robust and independent probe of the total density, free from many of the complications inherent in light-based modeling. A multitude of observational studies have been conducted exploring the connection between the lensing properties of galaxies and their environment and evolution \citep{treu2006,auger2008,treu2009,koopmans2009b,barnabe2009,barnabe2011,sonnenfeld2012}.  Simulation-based studies have shown that strong \revisions{gravitational} lensing observables are indeed sensitive to differences in baryonic processes \revisions{that} shape the mass distribution in lens galaxies  \citep{duffy2010,remus2017,peirani2017,mukherjee2021}. 

Such studies (both simulation-based and observational) have typically assumed a simple power-law ellipsoid mass distribution (PEMD;  \citealt{keeton1998,treu2010}).  This density profile has been remarkably successful at fitting observed properties of gravitational lens systems, given its simplicity.  It has been observed that dark matter and baryons together tend to form nearly isothermal profiles, a phenomenon \revisions{that} has been dubbed the ``bulge-halo conspiracy''  \citep{koopmans2009b,auger2010,dutton2014,xu2016}.  Realistically, however, the absence of additional radial and angular structure in lens mass models is overly simplistic, given the physical complexity of galaxy formation and evolution \citep{tollet2016}.  This complicates the interpretation of strong \revisions{gravitational} lens analyses \revisions{that} assume a PEMD. \cite{xu2017}, \cite{enzi2020}, and \cite{kochanek2020} all identify problematic biases that can arise from this simplifying assumption.  

Recent \revisions{applications of} gravitational lensing have required more complex mass models. In the field of time-delay cosmography, it has become standard practice to form a composite model containing both a baryonic and dark matter component \citep[e.g.,][]{wong2017,nightingale2019,rusu2020}, and to include environmental and kinematic information \citep[e.g.,][]{rusu2017,sluse2017,birrer2019,tihh2020,birrer2021,yil2021}. It is also becoming increasingly apparent that lens models should include azimuthal degrees of freedom as well \citep{kochanek2021,cao2022}.  Overly simplistic models have also been found to be problematic for detecting dark matter \revisions{sub-haloes} using flux-ratio anomalies, as the presence of disks and other baryonic structures can bias the results if not properly accounted for \citep{gilman2017,hsueh2016,hsueh2017,hsueh2018b,he2022}.

In this \revisions{paper}, we present the first study focused on probing the mass structure of a lens galaxy using \revisions{extended gravitational arcs observed at milli-arcsecond (mas) resolution with} very long baseline interferometry (VLBI). We consider several different parameterizations for the mass distribution \revisions{that} contain varying degrees of complexity, \revisions{and carry} out an objective Bayesian comparison between them.  Key to this analysis is the \revisions{extremely} high angular resolution afforded by global VLBI data, from which we jointly infer the lens parameters and pixellated source reconstructions using the method presented \revisions{by} \cite{powell2021}. We first describe the observations used in this work in Sections \ref{sec:data} and \ref{sec:aodata}. The lens model parameterizations are presented in Section \ref{sec:lensmodel}. We then review the Bayesian inference method in Section \ref{sec:inference}. We present the results, including an evidence-based comparison between lens models in Section \ref{sec:results}, with an interpretation of our results in Section \ref{sec:discussion}. Throughout this work, we use the \cite{planck2016} cosmology, with $H_0 = 67.8~\mathrm{km\,s^{-1}\,Mpc}$, $\Omega_m = 0.307$, and $\Omega_\Lambda = 0.693$.

\section{Radio interferometric data} \label{sec:data}

 \begin{figure*}
 \centering
   \includegraphics[scale=\figscale]{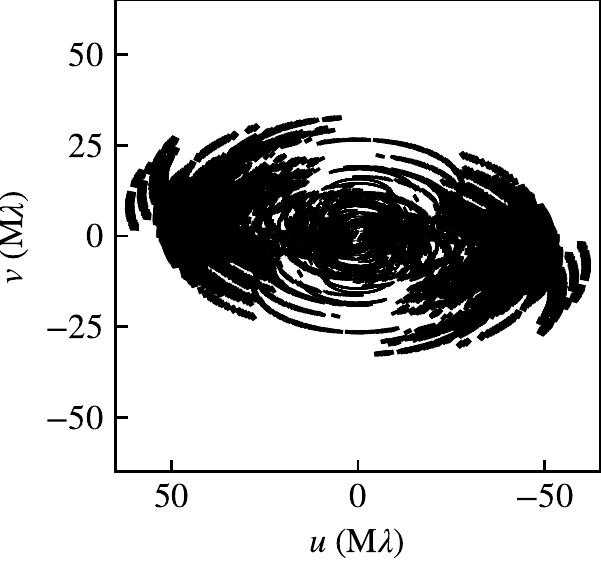}\hspace{0.7cm}
   \includegraphics[scale=\figscale]{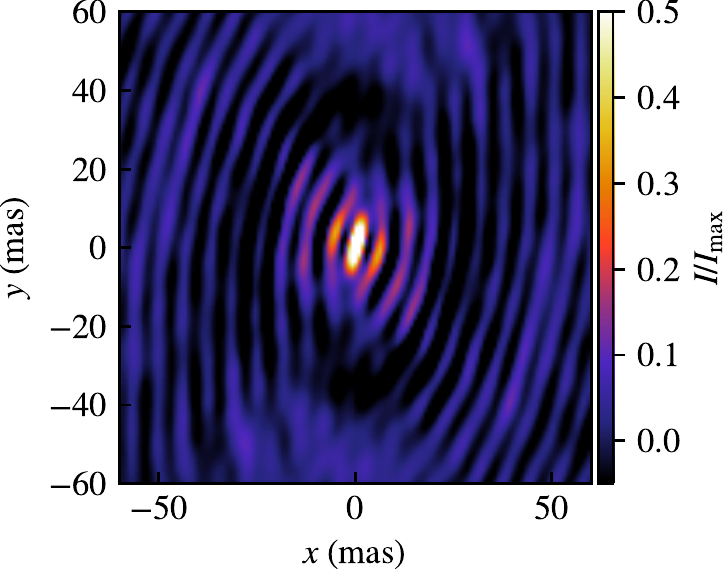}\hspace{0.5cm}
   \includegraphics[scale=\figscale]{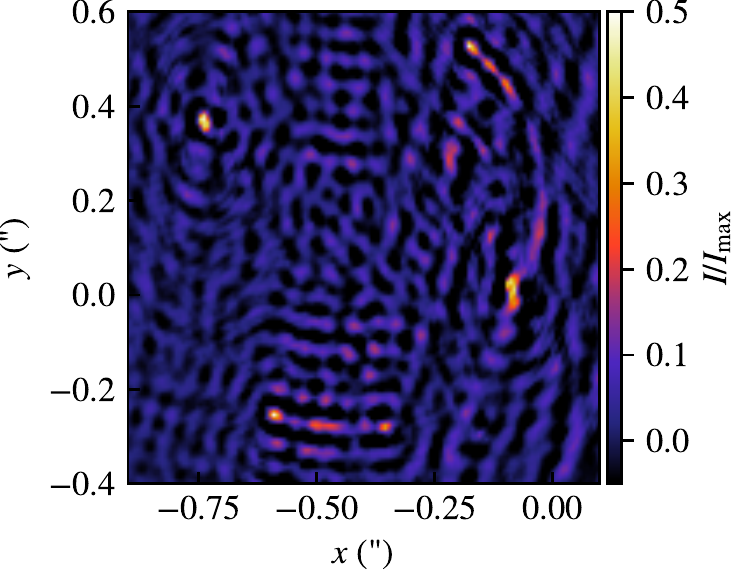}
  \caption{$uv$-coverage (left), naturally-weighted dirty beam (centre), and dirty image (right) of the global VLBI observation of \jzsfo. The main lobe of the dirty beam is $5.5\times1.8~\mathrm{mas}^2$ (FWHM) with a position angle of $9.8$ degrees.  The $(x,y)$ coordinates of the dirty image are given in arcseconds relative to the phase centre.  }
  \label{fig:data}
\end{figure*}

\jzsfo is a strongly-lensed quasar initially observed in the MIT-Green Bank Very Large Array (MG-VLA) survey \citep{lawrence1986}. Its discovery was presented along with follow-up observations at systematically higher angular resolution with the VLA and the Multi-Element Radio Linked Interferometer Network (MERLIN), as well as a rudimentary lens model by \cite{lehar1997}. \cite{alloin2007} published a study focusing on the dust and molecular gas content of the \revisions{background quasar}. They also proposed an improved lens model incorporating the mass of additional nearby galaxies, which were spectrosopically confirmed by \cite{tonry1999} to be members of a common group at $z_l=0.35$, as well as the source redshift of $z_s=3.2$.  Most recently, \cite{spingola2018} presented an analysis of a global VLBI observation of \jzsfo , while its high angular resolution optical, near-infrared and CO (1--0) properties were presented \revisions{by}  \citet{spingola2020}. \revisions{The radio} source is bright also at low frequencies and it has been observed using the long baselines of the Low-Frequency Array (LOFAR; \citealt{badole2022}). The global VLBI observations \revisions{of \citet{spingola2018}} are currently the highest angular resolution observations of any gravitational lens system containing extended \revisions{gravitational} arcs, with sharply resolved arcs and images localized to within a \revisions{a fraction} of a milli-arcsecond. In \revisions{Fig.}~\ref{fig:data}, we show the $uv$-coverage, dirty beam, and dirty image for this observation.

\subsection{Measurement sets and flagging} \label{sec:dataprep}

The observation of \jzsfo analyzed here was carried out on \revisions{2012 October 12} using \revisions{a} global VLBI array composed of 24 antennas from the European VLBI Network (EVN), the Very Long Baseline  Array (VLBA), \revisions{and the Green Bank Telescope} (project GM070; PI: McKean).  The total time on-source was 18.5 hours, with a visibility integration time of 2 s. The total bandwidth \revisions{was} 64 MHz, centred around 1.65 GHz. This \revisions{total} bandwidth \revisions{was} divided into 256 frequency channels (32 channels in each of 8 spectral windows). The calibration and data reduction was performed by \citet{spingola2018}, and we refer to their work for \revisions{further details}.

\revisions{From the dataset produced by} \citet{spingola2018}, we estimate the noise of the \revisions{visibilities} using the procedure described in Section~\ref{sec:noise}. We then flag all visibilities with \revisions{a} noise greater than 1 Jy \revisions{to remove any outliers}. Finally, we flagged the Effelsberg \revisions{to} Jodrell Bank baseline so as not to allow our inference to be dominated by this single, very sensitive baseline. The final calibrated and edited observation used in this work contains $2.5\times10^8$ unflagged visibilities.

\subsection{Noise estimation} \label{sec:noise}

The noise column provided in a \software{CASA} \citep{mcmullin2007} measurement set is computed from the radiometer equation, $\sigma \propto (\Delta \nu \Delta t)^{-\frac{1}{2}}$,
which depends on the channel bandwidth $\Delta \nu$ and the integration time $\Delta t$. However, this is a simple theoretical estimate \revisions{that} may not capture other instrumental and atmospheric effects that vary on timescales shorter than the full observation.  We instead measure the noise empirically from the data as follows.

We first partition the data by baseline, observation epoch, spectral window, and polarization. We further divide these data into 15-\revisions{min} blocks, giving $\sim 250$ visibilities per block. We then subtract time-adjacent visibilities from one another. Under the assumption that the sweep of each baseline across the $uv$ plane is sufficiently small between integrations, this difference between \revisions{neighbouring} visibilities cancels the sky signal and provides a sample of the noise. We then take the RMS of these time-differenced samples, corrected by $\sqrt{2}$ to account for the subtraction, to obtain our estimate of the noise. Using this differencing scheme, we attempt to utilize as much information from the data as possible by computing a detailed noise estimate for each visibility.

\subsection{Image plane} \label{sec:mask}

The image plane pixel scale and dimensions are chosen to meet two criteria. First, it must be large enough to contain all of the lensed light that we wish to model; we \revisions{choose} a 1.2 arcsec $\times$ 1.2 arcsec field of view. Second, the pixel scale must be small enough that the dirty beam is properly Nyquist sampled, which is determined by the $uv$ coverage of the observation. We accomplish this by choosing an image-plane resolution of $\npix=1024^2$ and a pixel size of 1.17 mas $\times$ 1.17 mas.

In order to aid the inference process, we mask the image plane (see \citealt{vegetti2008}). This serves both to reduce the dimensionality of the reconstructed source and to constrain the region of the image plane that is allowed to contain emission. This can be interpreted as a prior on the model. We have found from \revisions{simulations} that it is desirable to make the mask as tight as possible without excluding any real emission, as this helps to prevent the model from overfitting to the noise \revisions{(see \citealt{powell2021} for further details)}.  

To generate the mask, we use the \software{CLEAN}ed image of \jzsfo \revisions{produced} by \cite{spingola2018}. We first threshold the image at $5\sigma_\mathrm{RMS}$, where $\sigma_\mathrm{RMS}=41\mu\,\mathrm{Jy~beam^{-1}}$ is the residual RMS \revisions{map} noise. We then pad the resulting region of the image by three beams (3 mas $\times$ 9.8 mas) in all directions. \revisions{As} the emission comes from multiple disjoint components, we lastly connect these components along a path defined by the locations determined by \cite{spingola2018}. The resulting mask is shown in the upper-left panel of \revisions{Fig.}~\ref{fig:modelcmp}. The image plane (and hence the triangulated source grid) contains $\nsource=4.5\times10^4$ unmasked pixels.

\section{Keck adaptive optics data} \label{sec:aodata}

\begin{figure*}
 \centering
   \includegraphics[scale=\figscale]{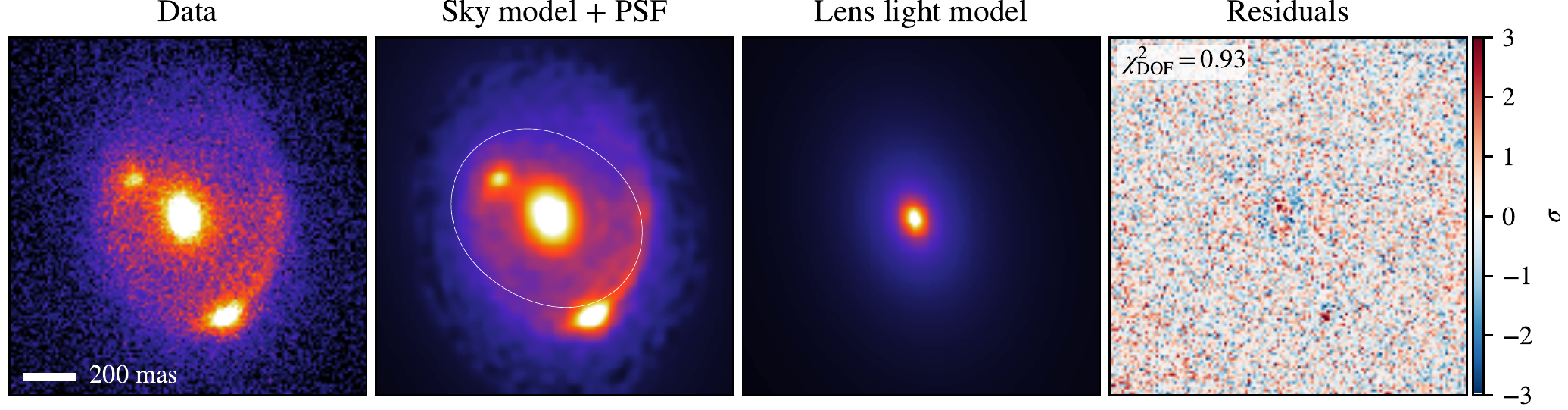}
  \caption{Light model for the lens galaxy, obtained from the Keck AO observation of \jzsfo (Section \ref{sec:aodata}). This model was obtained by jointly fitting a pixellated source surface brightness model with a \sersic profile for the lens light, following \protect\cite{ritondale2019}. We use this \sersic fit as a model for the baryonic mass content of the lens, assuming a constant mass-to-light ratio (Section \ref{sec:sersic}).  We show the source- and lens-plane surface brightness models for this observation along with the radio emission in \revisions{Fig.~\ref{fig:bestmodel}}.}
  \label{fig:keckmodel}
\end{figure*}

We also make use of an infrared (2.12 $\mu$m) observation of \jzsfo taken with the W. M. Keck-II
Telescope (Programs 2011B-U099 and 2012B-U079; PI: Fassnacht) as part of the Strong-lensing at High Angular Resolution
Programme (SHARP; e.g. \citealt{lagattuta2012}). The adaptive optics (AO) system on Keck provides a \revisions{point spread function (PSF)} with a \revisions{full width at half maximum (FWHM)} of the central part of about 65 mas. \revisions{A} detailed description of this observation and the data reduction process \revisions{used to produced the calibrated image} \revisions{(see Fig.~\ref{fig:keckmodel})} is \revisions{presented by \cite{spingola2020}}. We use this observation to extract a \sersic model for the lens galaxy light, as follows.  For the lens mass, we use a PEMD fit to the radio observation (Section \ref{sec:pemd}). We fix all lens parameters except for the position, which accounts for the loss of absolute position information during the phase-calibration process. While the PEMD model is not a perfect fit to the radio data, it is more than sufficient for this relatively low-resolution data. We then fit the lens position jointly with a pixellated source model and a \sersic profile for the light, following, \revisions{for example,} \cite{ritondale2019}.  We use a single \sersic profile, as we found a double-\sersic to be highly degenerate without significantly improving the fit.  We show the result of this fit \revisions{also} in \revisions{Fig.~\ref{fig:keckmodel}}.  We use this \sersic fit as a proxy for the stellar mass density profile, assuming a constant mass-to-light ratio (Section \ref{sec:sersic}).

\section{Lens models} \label{sec:lensmodel}

\begin{table*} 
\caption{Lens models compared in this work, as well as their components. Only the bottom 7 rows of this table are modeled in their own right; SR, MP, FG, and EP are considered only as components of the composite lens models.}
\centering
\begin{tabular}{l l l l}
\hline
\noalign{\vskip 0.1cm}
$\hyp$ & Description & Free parameters & Section\\
\noalign{\vskip 0.05cm}
\hline
\noalign{\vskip 0.1cm}
SR & \sersic profile & 1 & \ref{sec:sersic}\\
MP & Multipoles with $m=3,4$ & 4 & \ref{sec:mpoles}\\
FG$_\mathrm{fixed}$ & Field galaxies with masses fixed to values from \cite{spingola2018} & 0 & \ref{sec:model2}\\
FG$_\mathrm{free}$ & Field galaxies with free masses & 5 & \ref{sec:model2}\\
EP & External potential to 3$^\mathrm{rd}$ order & 4 & \ref{sec:expot}\\
\hline 
\noalign{\vskip 0.05cm}
PL & Power-law ellipsoid (PEMD) with external shear & 8 & \ref{sec:pemd}\\
PL+FG$_\mathrm{fixed}$ & PL with fixed field galaxy masses & 8 & \\
PL+FG$_\mathrm{free}$ & PL with free field galaxy masses & 13 & \\
PL+EP & PL with external potential & 12 & \\
PL+SR+EP & PL dark matter halo, \sersic stellar mass profile, and external potential & 13 & \\
PL+MP+EP & PL with multipoles and external potential & 16 & \\
PL+MP+SR+EP & PL dark matter halo with multipoles, \sersic stellar mass profile, and external potential & 17 & \\
\noalign{\vskip 0.05cm}
\hline 
\end{tabular} 
\label{tab:lensmodels}
\end{table*}

The lens mass profile of \jzsfo has been studied in detail by several authors to date. \cite{lehar1997} used VLA and MERLIN data (with a best resolution of 50 mas) to fit a lens model containing ellipsoidal power-law potentials for the main lens galaxy, plus four additional group galaxies. \cite{alloin2007} \revisions{used} the same data, along with {\it Hubble Space Telescope} ({\it HST}) observations at $\gtrsim80$ mas resolution from the CfA-Arizona Space Telescope LEns Survey \citep[CASTLES;][]{castles1999}, to model the lens and group using elliptical power-law density profiles. In addition, \cite{alloin2007} \revisions{included} a common group dark matter halo, claiming an improved fit. \cite{spingola2018} provided \revisions{an} improved lens \revisions{model} for this system using the same  global VLBI observation used in this work (Section \ref{sec:data}). They \revisions{proposed} both a single PEMD model, as well as a model \revisions{that included} additional PEMDs for five nearby group galaxies. Both of these models were only able to account for the observed image positions to within \revisions{$\sim 3$ mas}, which is quite significant considering the resolution and sensitivity of the observation. All of these aforementioned studies fit lens models by first identifying the image positions obtained from a separate imaging step, then fitting the lens parameters \revisions{that} best reproduce those positions. 

In this section, we enumerate an extended set of lens mass model parameterizations with which we attempt to improve upon previous modeling attempts using the method \revisions{presented by} \cite{powell2021}. We consider a set of models \revisions{that} allow for changes in both the angular and radial structure of the lens galaxy. We describe them in terms of their projected surface mass density $\kappa$, in units of the critical density $\Sigma_c$. The models as defined here lie their local reference frame, centred at the origin and aligned with the $x$ and $y$ axes.  In practice, we translate and rotate them to the lens centre $(x_0,y_0)$ and position angle $\theta_q$. Additionally, we include for all models an external shear component defined by its strength $\Gamma$ and direction $\theta_\Gamma$. We summarize these models in Table \ref{tab:lensmodels}.
We will compare them in Section \ref{sec:results} using their relative \revisions{log-evidence} (Section \ref{sec:evidence}). 

 \begin{figure*}
 \centering
   \includegraphics[scale=\figscale]{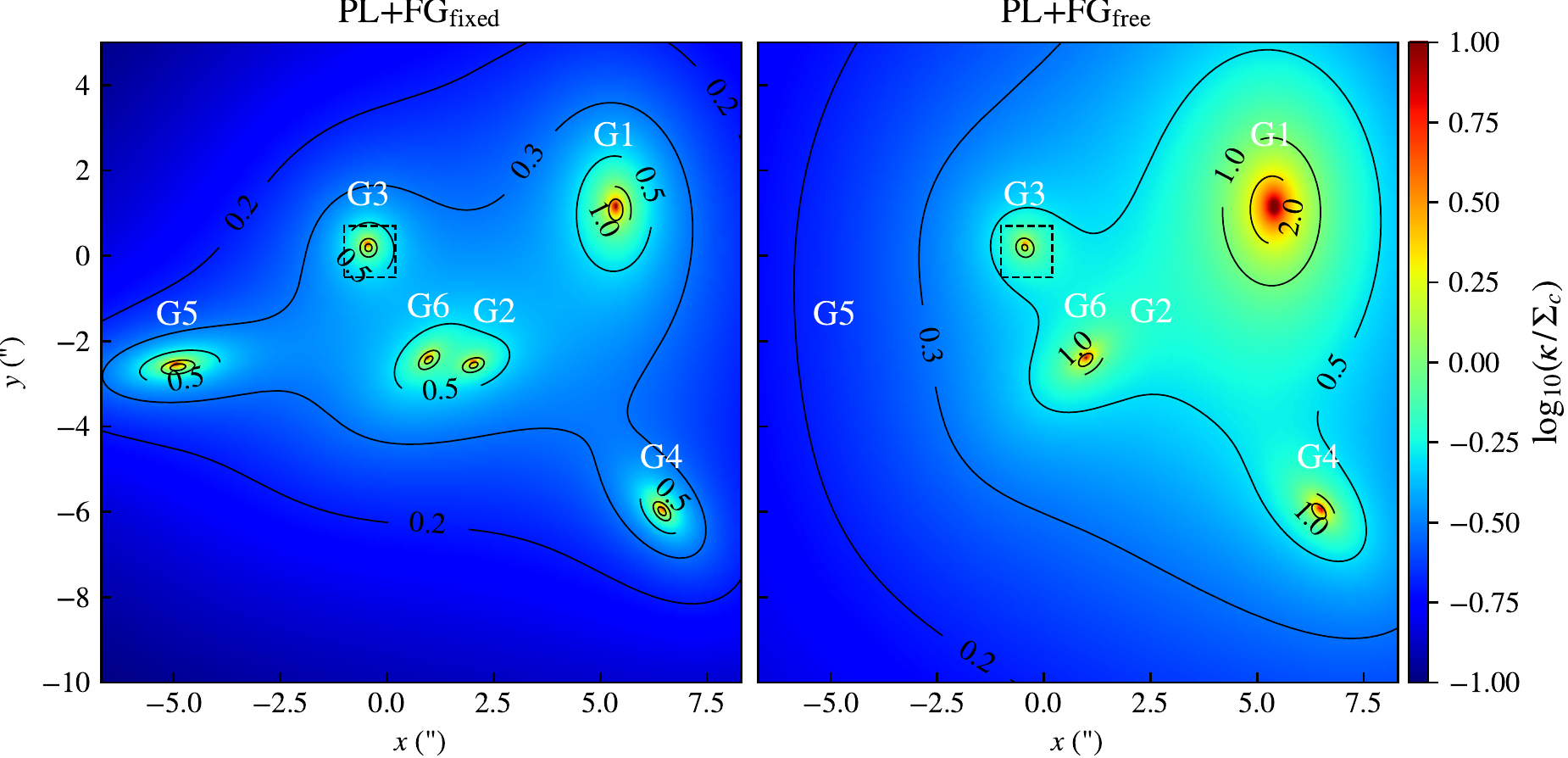}  
   \caption{Convergence maps showing the projected mass density (in units of critical density $\Sigma_c$) in a 15 arcsec $\times$ 15 arcsec field of view for model PL+FG (Section \protect\ref{sec:model2}), containing five additional group galaxies near the lens. We label them following \protect\cite{lehar1997,alloin2007,spingola2018}. The main lens galaxy G3 is shown enclosed in the black dashed square, which corresponds to the 1.2 arcsec $\protect\times$ 1.2 arcsec field used in our modelling procedure (see e.g. \revisions{Fig.}~\protect\ref{fig:modelcmp}). The left panel shows the model with group galaxy masses fixed to those from \protect\citet{spingola2018}, while the right panel shows the resulting convergence map when the galaxy masses are allowed to vary freely (see Section \protect\ref{sec:conv_mag_results}).  \revisions{As} the galaxy masses and global dark matter halo properties are not well-constrained by existing observations, we instead opt for a more generic expansion of the external potential around G3 (Section \protect\ref{sec:expot}) in order to capture environmental effects on the lens.} 
   \label{fig:model2}
\end{figure*}

\subsection{Power-law ellipsoid} 
\label{sec:pemd}

The power-law ellipsoid mass distribution (PEMD, see e.g. \citealt{keeton2001}) is a ubiquitous lens mass profile due to its simplicity and ability to fit a wide range of observed lens systems.  As this is the simplest model, with only 8 free parameters, we use the PEMD as our fiducial density profile, which we label PL.  

The PEMD has a normalized projected mass density
\begin{equation} \label{eq:spemd}
    \kappa(x,y) = \frac{\kappa_0\left(2-\frac{\gamma}{2}\right)q^{\gamma-\frac{3}{2}}}{2\left[ q^2 x^2  + y^2  \right]^{\frac{\gamma-1}{2}}}\,,
\end{equation}
where $\kappa_0$ is the mass normalization, $q$ is the elliptical axis ratio, and $\gamma$ is the power-law slope (with $\gamma=2$ corresponding to an isothermal power-law). In practice, we use the FASTELL library \citep{barkana1999} to compute the deflection angles. 

\revisions{For} composite models \revisions{that} contain a \sersic profile representing a baryonic mass component (Section \ref{sec:sersic}), we interpret the PL as a dark matter profile. While it is common practice to use an NFW profile for the dark matter in composite lens modeling \citep[e.g.][]{dutton2014,wong2017,rusu2020}, we allow for a free dark matter density slope. This choice is motivated by the fact that the dark matter profile in the inner $\sim$kpc of an elliptical galaxy is still poorly \revisions{understood}. Furthermore, in the limit where $r \ll r_s$, an NFW profile can be approximated by an $r^{-1}$ profile plus a mass sheet transformation. Typical scale radii for NFW haloes fitted to massive elliptical lens galaxies are on the order of $r_s \approx 10$ arcsec \citep{wong2017,rusu2020}, a factor of $\sim10$ \revisions{to 20} larger than the Einstein radii of their lens systems.

\subsection{\sersic profile}
\label{sec:sersic}

We form a composite mass model by combining PL (Section \ref{sec:pemd}) and a \sersic component.  In this context, the PL models a dark matter halo with a free inner density slope, while the \sersic profile models the baryonic mass content of the lens.  We label the \sersic model component as SR. The functional form of this profile is
\begin{equation} \label{eq:ser}
\kappa(\psi)=M_s \exp\left\{ -b_n \left[ \left( \frac{\psi}{R_s}\right) ^\frac{1}{n_s} -1\right] \right\},
\end{equation}
which we express in terms of the elliptical radius $\psi^2 \equiv q_s^2 x^2 + y^2$. $M_s$ is the total mass normalization, $R_s$ is the effective radius, and $n_s$ is the \sersic index. $b_n$ is a constant computed such that $M_s$ is the total mass of the profile. \revisions{
As equation \eqref{eq:ser} does not admit analytic expressions for the deflection angles, we compute them by numerically integrating the expressions for general elliptical profiles, \revisions{as derived by} \cite{keeton2001}.}

We fix $R_s=5.94$~\revisions{arcsec}, $n_s=6.30$, the position $(x_s,y_s)=(-0.422'',0.167'')$, position angle $\theta_s=15.4$~\revisions{deg}, and axis ratio $q_s=0.717$ to the best values obtained by fitting the Keck adaptive optics observation of this lens system (Section \ref{sec:aodata}; \revisions{Fig.~\ref{fig:keckmodel}}). The only free parameter in the SR mass profile is $M_s$, which is a proxy for the baryonic mass of the lens galaxy. For simplicity, we assume a constant mass-to-light ratio, as further information would be required to constrain spatial variations of this quantity.

\revisions{We note that the fit to the Keck AO data yields an unusually steep \sersic index of $n_s=6.30$. As a check of the robustness of our results to $n_s$, we repeated our analysis of the composite lens models using a de Vaucouleurs profile ($n_s\equiv4$), which is typical for massive elliptical galaxies. We found that enforcing this shallower slope for the baryonic component of the lens model does not significantly affect the inferred slope of the dark matter component, nor does it change the overall model ranking (Section \ref{sec:results}). However, the inferred total mass of the \sersic component, $M_s$, decreases by a factor of $\sim3.5$ when we fix $n_s\equiv4$. Given our simplifying assumption of a constant mass-to-light ratio, and the absence of absolute flux calibration and/or kinematic information, we cannot reliably constrain $M_s$ anyway (Section \ref{sec:stellarmass}).}

\subsection{Internal multipoles} 
\label{sec:mpoles}

As an extension to the elliptical PL profile, we include multipole-like terms describing internal angular structure in the mass distribution of the lens galaxy. This model is meant to encompass generic smooth deviations from ellipticity, which may arise from mergers, tidal forces, and/or baryonic processes (Section \ref{sec:intro}). The functional form of the convergence is
\begin{equation}
    \kappa_m(r,\theta) = r^{-(\gamma-1)}\left[a_m \sin(m\theta)+b_m\cos(m\theta)\right].
\end{equation}
Here, we express the convergence more naturally in polar coordinates, with $r$ in \revisions{arcsec}. $a_m$ and $b_m$ together describe the strength and orientation of the multipole perturbation. These coefficients give the strength of the density perturbation in units of the critical density $\Sigma_c$ at a radius of 1~\revisions{arcsec} from the lens centre. The slope $\gamma$ is fixed to that of the underlying PL (equation \ref{eq:spemd}). The potential and deflection angles are obtained trivially via the Poisson equation. 

We impose a Gaussian prior of width $\sigma=0.01$ on $a_m$ and $b_m$. Our choice of prior is motivated by \cite{kochanek2004}, who note a typical amplitude of $\kappa_0 \sqrt{a_4^2+b_4^2} \sim 0.005$ from numerical simulations. The conversion to our units is not exact, as they assume an isothermal ($\gamma = 2$) density slope, but it is sufficient for our purposes. We will see in Section \ref{sec:results} that this choice of prior is indeed able to accommodate the multipole amplitudes favored by the data. We include multipole perturbations up to order $m=4$, labelling this model MP. In the presence of both PL and SR model components, the position and slope of the multipoles are tied to the PL.  We do not consider multipole components with $m>4$ in this paper, as we wish to minimize potential degeneracy with \revisions{sub-haloes} in the lens \citep{evans2003, congdon2005}, \revisions{or along the line of sight to the background radio source}.

\subsection{Field galaxies} 
\label{sec:model2}

\cite{momcheva2006} identify a total of 13 galaxies within 15~\revisions{arcsec} of the main lens galaxy, which were spectroscopically confirmed to be members of the same compact group. Both \cite{lehar1997} and \cite{alloin2007} \revisions{included} additional mass components for these galaxies in their lens models. \cite{alloin2007} also \revisions{included} a dark matter halo common to the group. Most recently, \cite{spingola2018} also modeled this system with group galaxy properties inferred as follows: positions, ellipticities, and position angles were measured using archival optical {\it HST} data (GO-7495; PI: Falco). \revisions{A singular isothermal ellipsoid mass distribution (SIE; e.g. \citealt{keeton2001}) was assumed for each galaxy, with mass normalizations set} relative to the main lens galaxy using their optical magnitudes and scaling relations appropriate to their Hubble types \citep[e.g. ][]{mckean2005,more2008}. However, they also find that the inclusion of a global dark matter halo as in \cite{alloin2007} does not afford a well-constrained position or mass, so they do not include it in the model. As such, we also forego a global dark matter halo in our FG model.

In order to test whether external differential shear due to tidal forces from \revisions{neighbouring} galaxies can improve upon the fiducial PL model, we also test a mass model \revisions{that} includes these field galaxies. We use the same positions and ellipticities as found by \cite{spingola2018}.   We hereafter label this model as FG.  We test two variants of this model: FG$_\mathrm{fixed}$, in which we fix the masses to those obtained by \cite{spingola2018}, and FG$_\mathrm{free}$, in which we treat the field galaxy masses as free parameters.  We plot the total convergence of the group, for both fixed and best-fit free masses, in a 15 arcsec $\times$ 15 arcsec field of view in \revisions{Fig.}~\ref{fig:model2}, along with labels for the galaxies consistent with the aforementioned previous works.

\subsection{External potential} 
\label{sec:expot}

The masses of nearby galaxies (Section \ref{sec:model2}) are set by dynamical scaling relations, which can be unreliable in a group environment \cite[e.g. ][]{focardi2012,pelliccia2019,perez2020}. Additionally, the location and scale of a group-scale dark matter halo is unconstrained \citep{spingola2018}. We accommodate this uncertainty by considering a more generic alternative model for the external potential. Expanding to third order around the main lens \citep[see][]{kochanek1991,bernstein1999,keeton2001}, we express this potential as
\begin{equation}
    \phi_\mathrm{ext}(r,\theta) = \frac{\Gamma r^2}{2} \cos 2(\theta-\theta_\Gamma) + \frac{\tau r^3}{4} \cos (\theta-\theta_\tau) + \frac{\delta r^3}{6} \cos 3(\theta-\theta_\delta).
\end{equation}
The lowest-order term is simply the external shear with strength $\Gamma$ and angle $\theta_\Gamma$. We emphasize that this external shear is already included in all of our lens models; we show it here for completeness.  The term \revisions{that is} proportional to $\tau$ corresponds to a gradient in the surface mass density with magnitude $\tau$ and direction $\theta_\tau$. The last term captures a gradient of the external shear with strength $\delta$ and direction $\theta_\delta$. We label this model EP, and consider it as a more flexible alternative to FG.

 \begin{figure*}
 \centering
   \includegraphics[scale=\figscale]{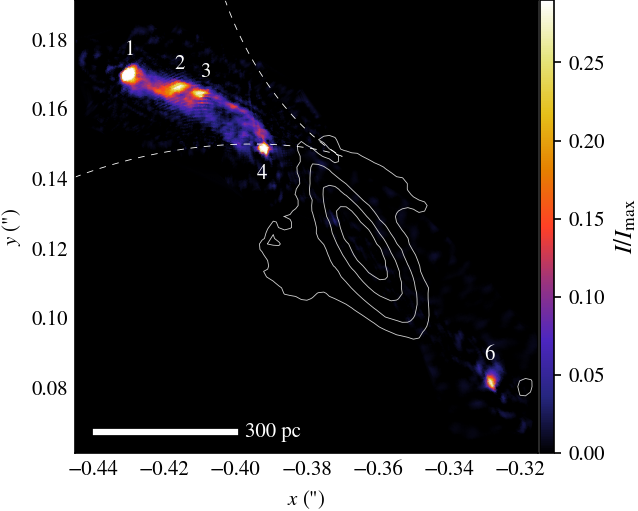}\hspace{0.8cm}
   \includegraphics[scale=\figscale]{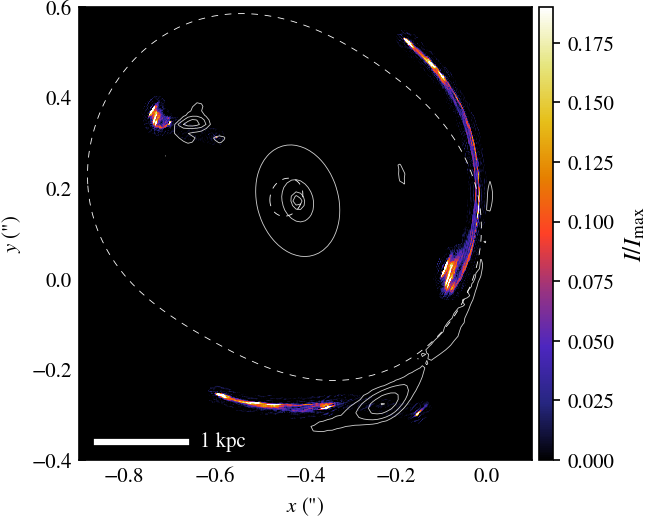}
  \caption{Maximum a posteriori source (left panel) and lens-plane (right panel) surface brightness reconstructions for PL+SR+MP, the best lens model in our evidence-based comparison (Section \ref{sec:evcomp}). The source is dominated by five distinct light components, \revisions{which for consistency, we label following} the numbering scheme of \protect\cite{lehar1997,spingola2018}. Caustics and critical curves are plotted as dashed white lines. The \revisions{colour} maps, which \revisions{are} normalized to the peak surface brightness, show the continuum radio emission, while the overlaid white contours show the source and sky emission reconstructed from the 2.12 $\mu$m Keck AO observation (Section \ref{sec:aodata}). We note that the apparent position angle of the reconstructed Keck AO source \revisions{is} biased by the strong magnification gradient in the direction perpendicular to the caustics.}
  \label{fig:bestmodel}
\end{figure*}

\section{Method} \label{sec:inference}

The Bayesian approach to jointly inferring the lens mass model and source surface brightness distribution has been well-established \citep{suyu2006,vegetti2008, hezaveh2016b, rizzo2018}. We carry out our analysis using a modified version of the visibility-space Bayesian gravitational lens modeling technique of \citet[for more details see also: \citealt{vegetti2008,rybak2015b, rizzo2018}]{powell2021}.  Here we review our notation and describe computational details specific to this work. 

\subsection{Bayesian inference} \label{sec:bayinf}

In radio interferometry, the data take the form of visibilities, which sample Fourier modes of the sky. Hence, our data $\data$ is a vector of $\nvis$ complex numbers.  The source $\source$ is a vector of length $\nsource$, which we represent on an adaptive Delaunay-tessellated grid as in \cite{vegetti2008}. The source light is mapped from the source plane to the image plane by the lens operator $\lensop(\etalens)$, which has dimensions $\nsource\times\npix$. The image dimension $\npix$ is set by the field of view and the angular resolution of the instrument, while $\nsource$ is determined by a light mask in the image plane (see Section \ref{sec:mask}). $\etalens$ is the set of parameters describing the lens mass distribution used to generate $\lensop$, where $\hyp$ denotes model parameterization (Section \ref{sec:lensmodel}). The instrumental response is $\dft$, the Fourier transform corresponding to the $uv$ coverage of the interferometer. Written as a matrix, $\dft$ is dense with dimensions $\nvis \times \npix$.   We assume additive Gaussian noise $\noise$ with covariance $\noisecov$.
With this notation, our model \revisions{$\model$} for the data $\data$ is 
\begin{equation}\label{eq:fmodel}
\revisions{\model} = \dft \lensop(\etalens) \source + \noise.
\end{equation}
We jointly infer $\source$ and $\etalens$ as follows. In the first level of inference, we compute the maximum a posteriori source $\smp$ for a given set of lens parameters $\etalens$ and source regularization strength $\lams$ as follows:
\begin{equation} \label{eqn:leastsquares}
   \msol \, \smp = (\dft\lensop)^T \noisecov \data\,,
\end{equation}
where
\begin{equation} \label{eq:msol}
    \msol \equiv \left[(\dft\lensop)^T\noisecov \dft\lensop + \lams \regul^T \regul \right]\,.
\end{equation}
 We solve equation \eqref{eqn:leastsquares} using a preconditioned conjugate gradient solver, where the Fourier operator $\dft$ is implemented using a nonuniform fast Fourier transform (NUFFT). We refer the reader to \cite{powell2021} for further details on the method.
 
 The operator $\regul$ in equation \eqref{eqn:leastsquares} is a discrete gradient operator defined on the Delaunay mesh.  The Gaussian source prior has covariance $\lams \regul^T \regul$, which penalizes large surface brightness gradients in the reconstructed source. Our choice of this form of prior is motivated by the lens equation: we know that conservation of surface brightness must hold for every lensed image of the source, such that if points on the image plane de-project onto the same point on the source plane, they must have the same surface brightness.  In the case of our adaptive Delaunay source-plane mesh \citep{vegetti2008}, source-plane pixel brightnesses are essentially interleaved from two or more different locations on the image plane. Hence, in a correctly focused source model, the surface brightness at two adjacent mesh vertices in the source plane should be very similar.  A source regularization \revisions{that} penalizes gradients encourages such a lens model. $\lams$ is a hyper-parameter \revisions{that} sets the strength of the source prior, such that a focused lens model will allow a larger $\lams$. We discuss this interpretation of the regularization term further in the Section \ref{sec:evcomp}.

In the second level, we infer the lens parameters $\etalens$ and source hyperparameter $\lams$. The posterior is 
\begin{equation}\label{eq:bayev}
P(\etalens,\lams\mid\data) = \frac{ P(\data\mid
		\etalens,\lams) \, P(\etalens) \, P(\lams)}{ P(\data)}\,.
\end{equation}
We use a uniform prior $P(\etalens)$ and log-uniform \revisions{prior} $P(\lams)$. The posterior (which is the evidence from the source-inversion step) is
\begin{multline} \label{eq:baypos}
    	2\log{P(\data \mid \etalens, \lams)} = -\chi^2 - \lams \smp^T \regul^T \regul \, \smp - \logdet \msol \\
    	+ \logdet (\lams \regul^T \regul)  + \logdet (2\pi \noisecov)\,.
\end{multline}
This expression follows from the marginalization over all possible sources $\source$ when the noise and source prior are both Gaussian. As $\regul$ and $\noisecov$ are sparse, the terms containing them are easy to evaluate.  Computing $\logdet\msol$ is non-trivial; we approximate it using the preconditioner from the inference on $\smp$ as described \revisions{by} \cite{powell2021}. 

\subsection{Fast \texorpdfstring{$\chi^2$}{chi-squared}}
\label{sec:fastchi2}

The $\chi^2$ term, required by the posterior in equation \eqref{eq:baypos}, is
\begin{equation} \label{eq:chi2}
    \chi^2 = (\dft\lensop\smp-\data)^T\noisecov(\dft\lensop\smp-\data).
\end{equation}
We speed its evaluation as follows. We first expand the quadratic form into its individual terms,
\begin{equation*} 
    \chi^2 = \smp^T \lensop^T \dft^T \noisecov \dft\lensop\smp - 2 \smp^T \lensop^T \dft^T \noisecov \data + \data^T \noisecov \data.
\end{equation*}
We next observe that $\dft^T \noisecov \data$ is the naturally-weighted dirty image, which we denote $\dirtyim$.  Similarly, $\dft^T \noisecov \dft$ performs a convolution with the naturally-weighted dirty beam, which we carry out efficiently using an FFT \citep{powell2021}. We define $\imcov \equiv \dft^T \noisecov \dft$, where the tilde indicates that $\imcov$ is implemented as a function rather than an explicit dense matrix. The last term, $\data^T \noisecov \data$, is a constant \revisions{that} need only be evaluated once. Making these substitutions yields
\begin{equation}  \label{eq:fastchi2}
    \chi^2 = \smp^T \lensop^T \imcov \lensop\smp - 2 \smp^T \lensop^T \dirtyim + \data^T \noisecov \data.
\end{equation}

We have, thus, shown that the $\chi^2$ can be computed entirely in the dirty image and gridded $uv$-plane bases, without the need to explicitly enter the (extremely high-dimensional) visibility space. No information is lost between the visibility space and the dirty image plane, given that the latter is sub-Nyquist sampled.
After precomputing and storing the dirty image and beam, each evaluation consists of just one forward/backward FFT pair and a few sparse matrix multiplications.  

This fast method for evaluating the $\chi^2$ is crucial to the feasibility of our analysis, which would otherwise require an expensive de-gridding operation at every posterior evaluation. We emphasize that although we do not explicitly fit the data in the visibility space, our technique is equivalent to within numerical precision. We stress that this only holds for the dirty image, and not the clean image plane, where the de-convolution process can lead to both loss of information and the introduction of image artefacts.

\subsection{Evidence computation and model comparison} \label{sec:evidence}

The final step of inference is to compare the relative probability of each lens mass parameterization, $\hyp$, given the observed data. This is done using the Bayesian evidence, which is computed by marginalizing over the entire parameter space of $\etalens$, $\lams$, and $\source$:
\begin{equation}  \label{eq:evidence}
    P(\data\,|\,\hyp) = \int P(\data\,|\,\etalens,\lams)\,
	P(\etalens) \, P(\lams) \, \mathrm{d}\lams \, \mathrm{d}\eta\,.
\end{equation}
Note that the marginalization over $\source$ has already taken place in equation \eqref{eq:baypos}. 

This integral has no closed-form solution, so it must be computed numerically.   We accomplish this using the \software{MultiNest} algorithm \citep{feroz2009}, which samples the full posterior distribution in the parameter space, while also computing the total evidence. For practical purposes, we express the evidence in logarithmic units using the notation
\begin{equation}  \label{eq:logevdef}
    \logev_\hyp \equiv \log P(\data\,|\,\hyp).
\end{equation}
We can then compare models using the difference in log-evidence, $\Delta \logev$, between the two. In this context, by ``different models'' we mean different parameterizations of the lens mass distribution (Section \ref{sec:lensmodel}). The Bayesian evidence provides us with an objective means to compare the ability of different models to explain the data, while automatically penalizing unnecessarily complex models.

\section{Results} \label{sec:results}

\begin{table*} 
\caption{ Summary of the main quantitative results for each lens model. We present the Bayes factor $\Delta\logev_\hyp$ relative to the best model PL+MP+SR, along with the optimal source regularization strength $\lams$. The RMS fractional difference in convergence, $\sigma_{\Delta\kappa}$, relative to PL+MP+SR, is measured inside a masked region within 17 mas (3 beam widths) of the lensed images (see the second row of \revisions{Fig.}~\ref{fig:magcmp}). The fractional difference $f_{H_0}$ in the measurement of $H_0$ inferred using time-delay cosmography for each model is stated relative to PL+MP+SR. The maximum change in flux for the brightest part of the source (source component \revisions{1}; see \revisions{Figs.}~\ref{fig:modelcmp} \revisions{and} \ref{fig:magcmp}) in each model, $|\Delta\mu|_{\rm max}$, is given relative to the flux-weighted mean magnification. Although the projected surface mass densities depart from the PL+MP+SR profile by only a few per cent (RMS) within the mask, the effect on inferences made using time-delays or flux ratios can be substantial. $\gamma_\mathrm{PL}$ is the inferred three-dimensional power-law slope for the PL component of each lens model, where $\gamma=2$ is isothermal.  For composite PL+SR models, $\gamma_\mathrm{PL}$ represents the slope of the dark matter component. The last column gives the total mass of the baryonic \revisions{\sersic} component for the SR models.}
\centering
\begin{tabular}{l l l l l l l l l}
\hline
\noalign{\vskip 0.1cm}
$\hyp$ & $\Delta\logev_\hyp$ & $\lams$ ($\times 10^9$) & $\sigma_{\Delta\kappa}$ (\%) & $f_{H_0}$ (\%) & $|\Delta \mu|_\mathrm{max}$ (\%) & $\gamma_\mathrm{PL}$ & $\log_{10}(M_s/M_\odot)$  \\
\noalign{\vskip 0.05cm}
\hline
\noalign{\vskip 0.1cm}
PL+MP+EP & $\mathbf{\equiv 0}$ & $10.8$ & $\equiv 0$  & $\equiv 0$ & $\equiv 0$ & 1.87 & -\\
PL+MP+SR+EP & $\mathbf{-350}$ & $10.6$ & $1.0$  & $+4.8$ & $0.9$ & 1.82 & 11.10\\
PL+SR+EP & $\mathbf{-5975}$ & $7.3$ & $3.1$  & $+10.3$ & $14.5$ & 1.88 & 11.19\\
PL+EP & $\mathbf{-9327}$ & $6.0$ & $1.7$  & $-3.6$ & $20.8$ & 1.84 & -\\
PL+FG$_\mathrm{free}$ & $\mathbf{-9863}$ & $5.8$ & $5.8$  & $-57.7$ & $21.7$ & 1.76 & -\\
PL & $\mathbf{-14984}$ & $4.3$ & $3.3$  & $+3.2$ & $8.0$ & 1.90 & -\\
PL+FG$_\mathrm{fixed}$ & $\mathbf{-22043}$ & $2.8$ & $6.6$  & $-19.2$ & $12.4$ & 1.90 & -\\
\noalign{\vskip 0.05cm}
\hline
\end{tabular}
\label{tab:logev}
\end{table*}

\subsection{Source- and lens-plane surface brightness} \label{sec:source_results}

The best-performing lens model in our analysis is PL+MP+EP; we address the Bayesian model ranking in detail in Section \ref{sec:evcomp} and Table \ref{tab:logev}.  We show the MAP source and sky surface brightness reconstructions for this model in \revisions{Fig.}~\ref{fig:bestmodel}. In addition, we overlay contours corresponding to the (rest-frame) optical source and sky reconstructed \revisions{emission} from the Keck AO data (Section \ref{sec:aodata}) using the same lens model.
The double-jet structure is clearly visible, with several distinct hot spots.  To aid our discussion, we have numbered these light components 1, 2, 3, 4, and 6 (omitting 5 to remain consistent with \citealt{lehar1997} and \citealt{alloin2007}, who detect this component only at shorter wavelengths). 

The north-western jet, comprised of components 1 \revisions{to} 4, extends 75 mas (550 pc in projection at the source redshift $z_s=3.2$) from the centre of the host galaxy. The counter-jet is visible only as the relatively dim component 6, \revisions{which is} 50 mas (350 pc \revisions{in projection}) to the south-east of the galaxy.  The high surface brightness of the north-western jet relative to its counterpart indicates that it is relativistically beamed along the line-of-sight.  These results are consistent with the projected size of the jet obtained from the parametric lens model of \cite{spingola2018} and from the pixellated reconstruction of \revisions{37.8~GHz} VLA data \citep{spingola2020}. Recently, \citet{badole2022} found that low-frequency radio emission from the jet observed using \revisions{LOFAR} is extended on a similar size in projection, implying that the radio emission is all confined within the jet and there is no emission associated with any extended lobes.

A side-by-side comparison of surface brightness reconstructions for all lens models considered in this work is shown in \revisions{Fig.}~\ref{fig:modelcmp}. Each column of this figure corresponds to the maximum a posteriori (MAP) lens parameters, sky, and source for that model. The first row shows the lens-plane surface brightness map along with the critical curves. While the model residuals formally exist in the visibility space (equivalently, the dirty image space), in the second row we show a normalized image-plane representation of the residuals, which is computed as follows:
\begin{equation} \label{eq:imresid}
    \vec{r}_\mathrm{im} = \frac{1}{\sqrt{\nvis}} \dft^T \oper{C}^{-\frac{1}{2}} (\dft \lensop \source - \data).
\end{equation}
In a similar fashion to the fast $\chi^2$ evaluation (Section \ref{sec:fastchi2}), we can re-factor equation \eqref{eq:imresid} in a way that avoids explicit visibility-space computations.   The bottom two rows of \revisions{Fig.}~\ref{fig:modelcmp} show the source surface brightness for each model at different levels of detail.

\subsection{Bayesian model comparison} \label{sec:evcomp}

In order to objectively differentiate the ability of these models to explain the data, we now turn to the Bayesian evidence (Section \ref{sec:evidence}).  \revisions{Since} we care only about the \emph{relative} evidence for each model, we compare them in terms of the difference between their logarithmic \revisions{evidence}, which we denote \revisions{as} $\Delta \logev_\hyp$.  We summarize these results in Table \ref{tab:logev}. 

The best model is PL+MP+EP, containing an elliptical power-law, angular multipole perturbations and an external potential contribution.  We hereafter fix $\Delta \logev_\mathrm{PL+MP+EP} \equiv 0$, comparing the other models relative to this one.  The second-best model is PL+MP+SR+EP, with $\Delta \logev_\mathrm{PL+MP+SR+EP}= -350$.  While these models contain the most free parameters (16 and 17, respectively; see Table \ref{tab:lensmodels}), the Bayesian evidence also penalizes (via Occam's razor) superfluous degrees of freedom, as demonstrated by the fact that the PL+MP+SR+EP model is slightly disfavoured relative to PL+MP+EP. We therefore interpret this result as a truly data-driven preference for the presence of angular and radial structure in this lens system. The next-best models are PL+SR+EP ($\Delta \logev_\mathrm{PL+SR+EP}= -5975$) and PL+EP ($\Delta \logev_\mathrm{PL+EP}= -9327$). The simple elliptical power-law model gives $\Delta \logev_\mathrm{PL}= -14984$.  The lens models containing field galaxies (FG) are amongst the worst-performing ones, with $\Delta \logev_\mathrm{PL+FG_{\rm free}}= -9863$ and $\Delta \logev_\mathrm{PL+FG_{\rm fixed}}= -22043$. We address these results in detail in the discussion section.

\subsection{Source regularization and \texorpdfstring{$\chi^2$}{chi-squared}} \label{sec:sourcereg}

 \begin{figure}
 \centering
   \includegraphics[scale=\figscale]{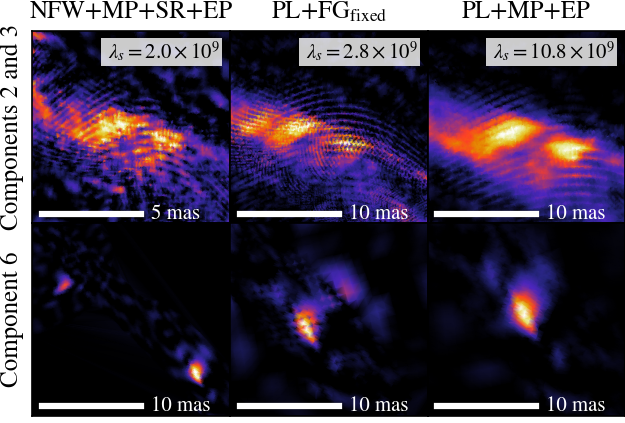}
  \caption{Extreme detail of the source surface brightness for the worst (middle column) and best (right column) lens models on two separate regions of the source (see \revisions{Fig.}~\ref{fig:bestmodel}). In the left column, we show the best-fit source for a model \revisions{that} includes a true NFW profile for the dark matter, rather than a power-law; we do not consider NFW+SR models in detail this paper, as they yield exceedingly poor fits to the data (see Section \ref{sec:densityslope}). The purpose of this figure is to illustrate the effect of a poorly-focused lens model on the source reconstruction. In the top row, we see Moir\'{e}-like stripes \revisions{that} occur when adjacent source-plane grid points are lensed forward to the wrong locations in separate images in the lens \revisions{plane}, leading to many large gradients on small scales in surface brightness; these stripes are most prominent in the worst model (PL+FG$_\mathrm{fixed}$), while NFW+MP+SR+EP is completely disrupted. In the best model, these features are present, but to a much lesser extent, and the distinct light components are clearly better captured. In the bottom row, we see that NFW+MP+SR+EP and PL+FG$_\mathrm{fixed}$ fail to align component 6, and the model attempts to fit the data by simply duplicating this feature on the source plane. The best model merges these into one coherent component. The ability of model PL+MP+EP to better focus the source suppresses large surface brightness gradients and prefers a larger prior strength $\lams$. See Sections \ref{sec:bayinf} and \ref{sec:sourcereg} for further discussion.}
  \label{fig:zoomcmp}
\end{figure}

In order to interpret the Bayesian model comparison in an intuitive way, let us consider the effects of the source prior and $\chi^2$ on our model comparison. The MAP model gives a reduced $\chi^2_\mathrm{DOF}=1.03$. The maximum fractional difference in the $\chi^2$ between any two of the models is $10^{-5}$, and the residual maps are indistinguishable by eye (\revisions{Fig.}~\ref{fig:modelcmp}).  This indicates that the model attempts to fit the data equally well (within the constraints of the source prior), regardless of the lens profile. Rather, the difference in log-evidence between the mass profiles is primarily driven by the ability of the lens model to focus the source. Lens models \revisions{that} correctly align the back-projected images on the source plane are able to reconstruct a source in which the presence of large gradients is minimized. We illustrate this in \revisions{Fig.}~\ref{fig:zoomcmp}, where it can be clearly seen that the worst model, PL+FG, contains stripes of rapidly varying surface brightness, as well as multiple copies of component 6.  In the best model, PL+MP+SR, where the source is better (though still not perfectly) focused, these strong gradients on small scales are much less prominent. In \revisions{Fig.~\ref{fig:zoomcmp}}, we also show the reconstructed source for a lens model with an NFW profile for the dark matter distribution, a \revisions{\sersic} profile for the baryonic component, and the inclusion of multipoles and an external potential (NFW+MP+SR+EP) . It can be seen that this model completely fails to focus component 6, which splits \revisions{into} two peaks of surface brightness separated by more than 10 mas. We further discuss this model in Section \ref{sec:densityslope}, but otherwise ignore it \revisions{for} the rest of the paper, given its failure to fit the data appropriately.

\revisions{The defining feature of the adaptive Delaunay source plane discretization is that the surface brightness at each source-plane vertex maps to exactly one image-plane pixel. Hence, any correlation between source-plane pixels (including enforcement of the lens equation) \emph{must} be explicitly encoded in the source prior. This motivates} our choice of \revisions{a} gradient-penalizing source prior (Section \ref{sec:bayinf}). The effect is a preference towards lens models \revisions{that} produce a better-focused source, and which properly align pixels of similar surface brightness on the source plane.  Such models admit a stronger source regularisation via a larger optimal value for $\lams$. Hence, we interpret the preferred $\lams$ as a proxy for the goodness-of-fit of the lens profile; we give $\lams$ for each model in Table \ref{tab:logev} and in \revisions{Fig.}~\ref{fig:modelcmp}. Differences in the Bayesian evidence are primarily manifested in the source regularisation term, via the ability of a given lens model to correctly align the back-projected images on the source plane, in agreement with the lens equation and conservation of surface brightness. \revisions{In a simple test, we artificially varied $\lams$ between $2.8 \times 10^9$ and $1.08 \times 10^{10}$ for each lens model, keeping the lens parameters fixed to their MAP values.  We found that the resulting source surface brightness maps change only at the few per cent level for all lens models. This test confirms that the model ranking is primarily driven by the capability of each mass model to produce a well-focused source, rather than the source regularization strength $\lams$.}

\revisions{The effect of priors on data fitting and model ranking is central to Bayesian inference in general. The specific case of priors for pixellated source reconstructions in gravitational lensing is a subtlety, which has been studied in some detail by several authors to date (e.g. \citealt{suyu2006, galan2021, vernardos2022}). Although there exist a plethora of possible forms for the source prior, for this work, we restrict ourselves to the gradient-based prior due to its physical motivation by the lens equation, as discussed above.  An additional subtlety that can be interpreted as part of the source prior is the choice of image-plane mask, which determines the number of source degrees of freedom. The number of pixels within the mask is determined both by the $uv$-coverage of the observation (via the Nyquist sampling theorem), as well as the prior belief on the extent of the true sky emission; see Section \ref{sec:mask}.  While we expect the lens model ranking in this work to be robust to our choice of source prior, we reserve a detailed comparison of prior choices for future work.}

\revisions{We note that} even the most preferred lens model we test here still contains spurious features (on the scale of a few mas) caused by imperfect focusing of the source. Hence, for these relatively smooth parametric lens profiles, we are still in a regime where the source regularisation simply encourages the lens model to focus, rather than imposing some physical information on the source itself. In a follow-up paper, we will test whether the presence of low-mass haloes within the lens galaxy and along its line of sight can further improve our source reconstruction.

\subsection{Convergence, magnification, and lens parameters} \label{sec:conv_mag_results}

\begin{landscape}
 \begin{figure}
 \centering
   \includegraphics[scale=\figscale]{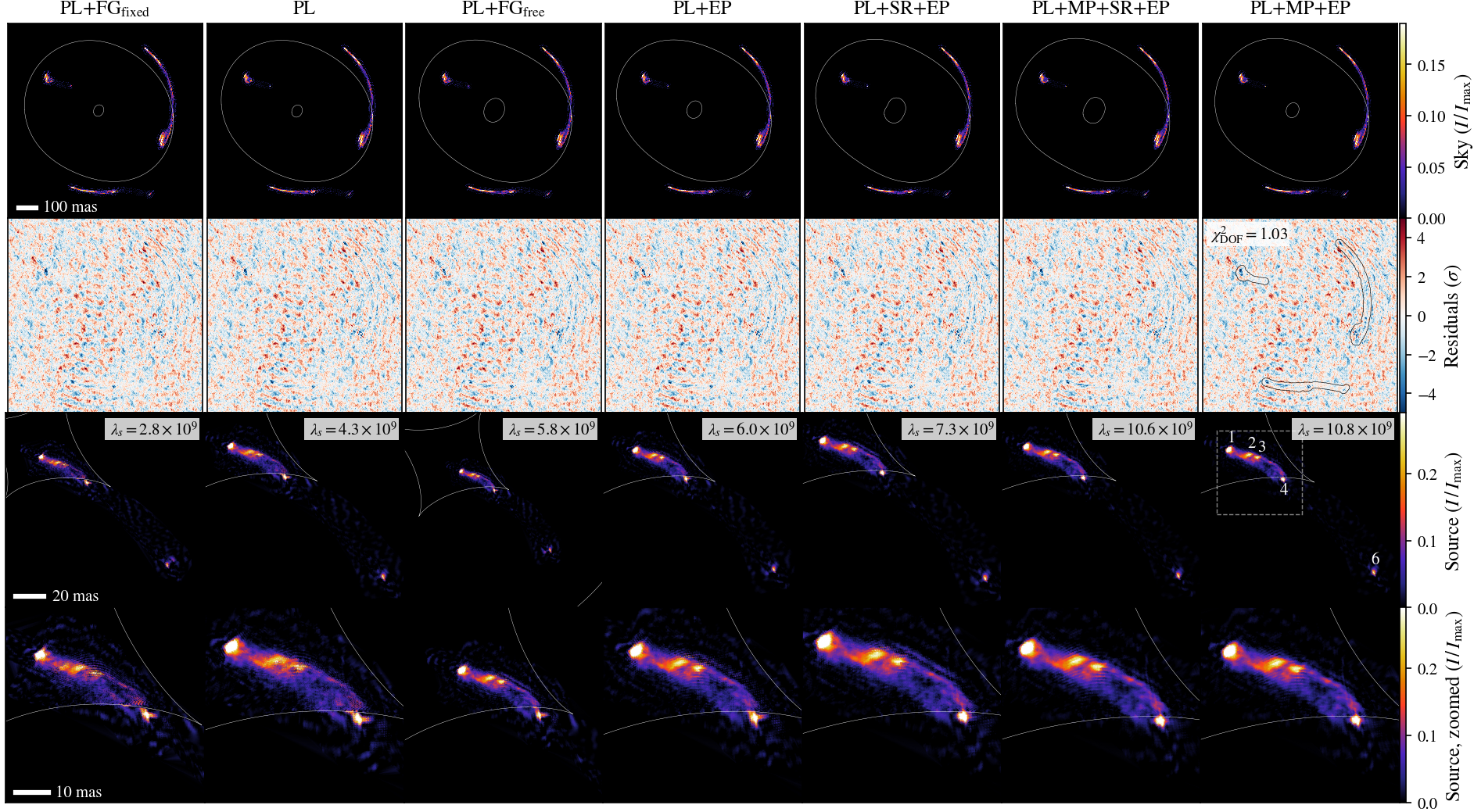}  \caption{Comparison of maximum a posteriori (\revisions{MAP}) source and sky models for each lens model, ordered from left to right by increasing Bayesian evidence. Colour scales and physical sizes are consistent across each row. 
   Top row: The lens-plane sky model, with critical curves in white. Second row: A normalized image-plane representation of the residuals (see equation \protect\eqref{eq:imresid}). The mask is shown as a thin black outline in the rightmost panel. The top two rows use the same physical extent and scale for each panel. Third row: The source surface brightness for each model, which are translated relative to one another depending on the mean deflection of the lens model. In the rightmost panel of the third row, we label the brightest light components of the source following \protect\cite{lehar1997,alloin2007,spingola2018}. Bottom row: A zoomed view of the main source features (with an extent shown by the dashed square in the right-hand panel of the third row) in order to highlight the ability of each lens model to focus the source. }
  \label{fig:modelcmp}
  \end{figure}
\end{landscape}

In \revisions{Fig.}~\ref{fig:modelcmp}, we compare the convergence and magnification properties between each lens model. The top row shows the total convergence maps.  A noteworthy \revisions{result} is that in the composite PL+SR models, the dark PL and baryonic SR components prefer not to share a common centroid (the SR position is fixed by the lens galaxy light; see Section \ref{sec:sersic}). Given the group environment of this lens, the presence of an offset between dark and baryonic components is not implausible; several weak-lensing studies of galaxies in groups and clusters provide evidence that light must not necessarily follow mass \citep{massey2011,george2012,foex2014,viola2015,massey2015}.

In the second row of Figure \ref{fig:modelcmp}, we compare the fractional differences in convergence relative to PL+MP+SR, defined as
\begin{equation*}
    \Delta\kappa\equiv(\kappa-\langle\kappa\rangle)/\kappa_B.
\end{equation*}
Here, $\kappa_B$ is the convergence of the best model. $\langle\cdot\rangle$ denotes an average of the convergence within the light mask (Section \ref{sec:mask}); subtracting this mean convergence removes the mass-sheet degeneracy from the comparison. We summarize the difference in convergence from the best model using $\sigma_{\Delta \kappa}$, which is the RMS of $\Delta \kappa$ within the mask. Composite FG models have the largest departure in convergence from the best model, with $\sigma_{\Delta \kappa}\approx 5$ \revisions{to 6} per cent. The convergence in all other models is very close to that of PL+MP+EP, with a maximum $\sigma_{\Delta \kappa} = 3.3$ per cent in model PL.  These results are summarized in Table \ref{tab:logev}.

In the third row of \revisions{Fig.}~\ref{fig:modelcmp}, we show magnification maps for each lens model. We also label the magnifications of the four images of the brightest source component 1. The last row shows maps of the fractional differences in magnification relative to PL+MP+EP, rescaled by the flux-weighted mean magnification. We additionally show the fractional change in magnification at the four brightest images. These magnifications vary on the $\sim10$ per cent level (Table \ref{tab:logev}). 

The numerical values of the inferred lens parameters for all parameterizations are listed in Table \ref{tab:lenspars}. We also present cornerplots of the full posterior distributions in Appendix \ref{app:lenspars}. The lens parameters are extremely well-constrained by the data. The source consists of multiple compact light components arranged along a  $\sim125~\mathrm{mas}$ span (\revisions{Fig.}~\ref{fig:bestmodel}), which in turn are projected onto different radii and angular positions in the lens plane. This lever-arm geometry, along with the high angular resolution of the data, results in unprecedented constraining power on the lens mass distribution. However, we warn the reader that the quoted errors are somewhat under-estimated. It is a well known fact that \software{MultiNest} tends to return overly-optimistic uncertainties. Moreover, \citet{nightingale15} have shown that having a deterministic relation between the Delaunay vertices on the source plane and the lens mass parameters, as it is the case here, can also lead to an under-estimation of the errors. To compensate for these effects, we follow \citet{rizzo2018} and also provide more realistic uncertainties by summing in quadrature the errors from \software{MultiNest} and the difference between the MAP lens parameters obtained by \software{MultiNest} and those obtained from a simple down-hill simplex optimisation.

 \begin{figure}
 \centering
   \includegraphics[scale=\figscale]{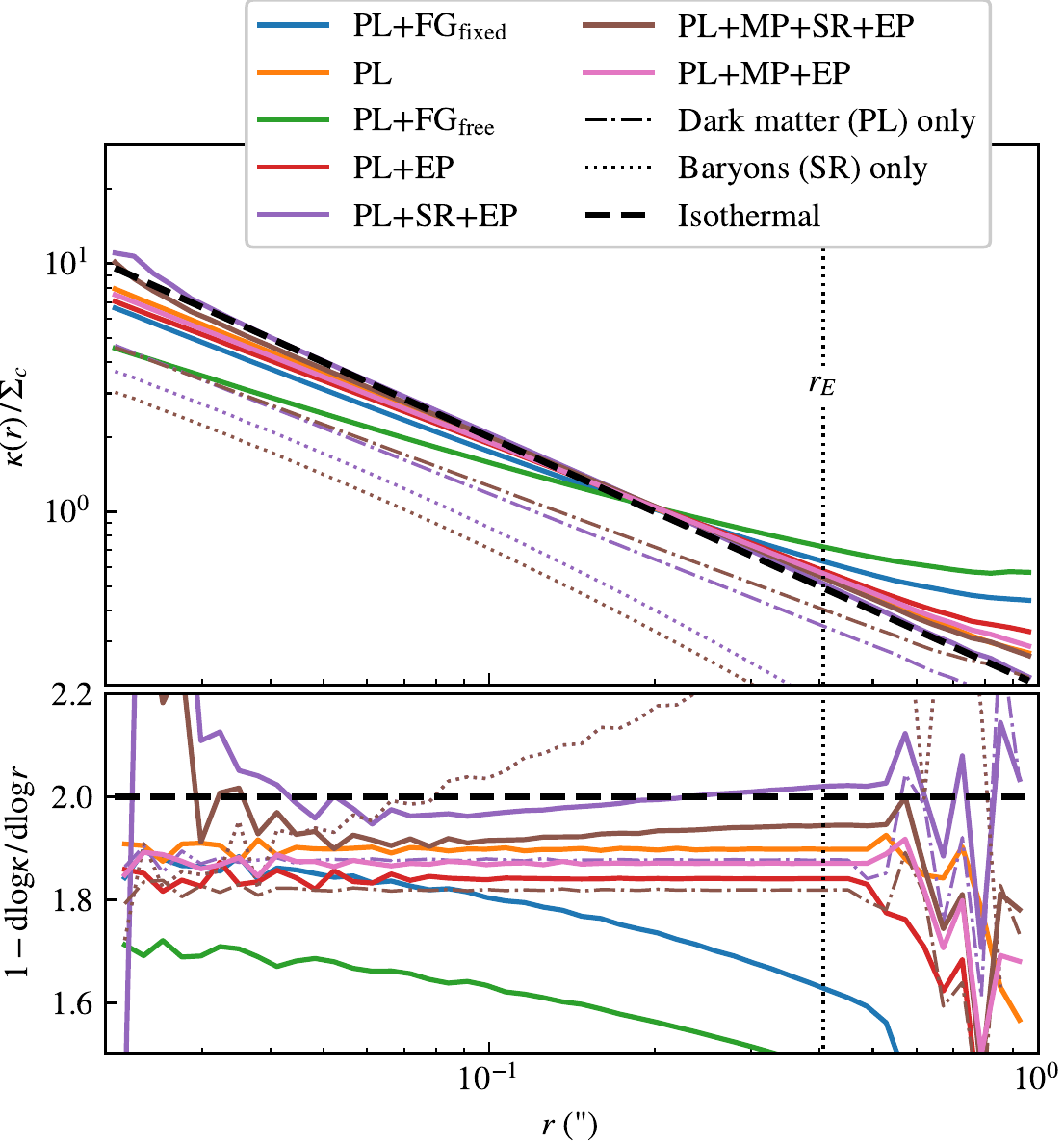}
  \caption{Azimuthally-averaged total surface mass density profiles for all lens models. We also show the PL- and SR-only profiles, for composite models which include both. In the bottom panel, we plot the corresponding logarithmic density slopes. The thick dashed line represents an isothermal ($\gamma=2$) profile, and the vertical dotted line shows the location of the Einstein radius.  The total density slopes inferred in this work are consistently sub-isothermal, while the dark-matter-only PL slopes from the composite models are significantly steeper than NFW (Section \ref{sec:densityslope}). }
  \label{fig:slopecmp}
\end{figure}

\section{Discussion} \label{sec:discussion}

\subsection{Lens mass distribution}

In this work, we have presented the first analysis of a lens system observed with VLBI at milli-arcsecond resolution using a pixellated source surface brightness model. While we consider models of varying degrees of complexity, we find that the simplest model PL focuses the source remarkably well, with deflection angle corrections only on the $\sim 5$ mas level needed to (almost) perfectly focus the source as in PL+MP+EP.  

\subsubsection{Mass density slope}
\label{sec:densityslope}

In \revisions{Fig.}~\ref{fig:slopecmp} we plot the total surface density profiles, measured by azimuthally-averaging the convergence maps shown in \revisions{Fig.}~\ref{fig:magcmp}, along with their logarithmic density slopes. Empirical density slopes for non-composite models match their parametrically-defined slopes, as expected. Models PL, PL+FG$_\mathrm{fixed}$, PL+EP, and PL+MP+EP, which contain no separate baryonic component, exhibit a total mass density power-law slope that is slightly shallower than isothermal. These are roughly consistent (at the $\sim1.5\sigma$ level) with slopes measured from the Sloan Lens ACS sample (SLACS; \citealt{auger2010}), which have a mean and scatter of $\gamma_\mathrm{SLACS}=2.078\pm0.16$.  Hence, it seems that the bulge-halo conspiracy \citep{koopmans2009b,auger2010,dutton2014,xu2016} lives on even for a lens system observed at \revisions{mas-scale} angular resolution.  In this respect, our analysis validates the use of a simple PEMD model for those applications of gravitational lensing where only the large-scale properties of the global mass model are relevant. This has important implications for modelling the large number of gravitational lenses to be found with, for example, {\it Euclid}, where the angular resolution is relatively low ($>100$ mas) and a simple PEMD will most likely be assumed. 

Our total slope values are also consistent, within the error, with the distribution of the SLACS, SL2S, and BELLS lenses as reported by \citet{mukherjee2021}. This result confirms that strong gravitational lens galaxies prefer galaxy formation models with weaker stellar and AGN feedback, which are, however, ruled out by other observations \citep{duffy2010, mukherjee2021, remus17, peirani18}.

Interestingly, the total density slopes for the composite models are closer to isothermal than any other models, with $\gamma_\mathrm{tot}=2.02$ and $1.94$ (measured at the Einstein radius) for models PL+SR+EP and PL+MP+SR+EP, respectively.  The effect of the mass sheet in the FG models is also clearly visible, with the slope declining consistently with radius.
Models with a baryonic component, that is, PL+MP+SR+EP and PL+SR+EP, have a dark matter mass density slope of $\gamma_{\rm PL} = 1.82$ and $\gamma_{\rm PL} = 1.88$, respectively. These values are significantly steeper than \revisions{the} inner slope of an NFW profile. While the exact slope values are likely to be affected by our assumption of a constant mass-to-light ratio, we notice that the NFW+MP+SR+EP model results in a significantly unfocused source as shown in \revisions{Fig.}~\ref{fig:zoomcmp}. While \cite{dutton2014} determine that the dark matter in gravitational lens systems is well-described by NFW profiles, this is an ensemble result \revisions{that} may include significant variation between galaxies, as well as redshift dependence. The slope of the inner dark matter profile is likely highly dependent on the cooling rates and levels of baryonic feedback for this specific galaxy \citep[e.g.][]{duffy2010}, as well as the dark model and its interplay with feedback \citep[e.g.][]{despali2019}. It may also be possible that an NFW+MP+SR+EP model with a varying mass-to-light ratio may result in a better fit to the data. More information is needed to relax our assumption on the mass-to-light ratio and test such a model.

\revisions{We finally note that, using the same VLBI data, \citet{spingola2018} inferred a density slope for this lens that is slightly steeper than isothermal. The fact that we obtain slightly different lens model parameters is not unexpected, as \citet{spingola2018} fit the model using the image-plane positions of a handful of discrete light components, which provides far fewer constraints than the full pixellated source surface brightness distribution. This demonstrates the additional constraining power that is contained in the highly resolved extended gravitational arcs. }

\subsubsection{Angular and radial structure}

We find strong evidence for the presence of both angular and radial structure beyond simple ellipticity in the mass distribution of this lens, with PL+MP+EP and PL+MP+SR+EP preferred over all other models. Based on the alignment of the multipoles relative to the major axis of the lens, the mass of this lens is neither ``boxy'' nor ``disky,'' but rather approximately halfway in-between. This may be due to tidal interactions between the main lens galaxy and other members of its group (G2 and G6 are close and lie in approximately the right direction); \cite{zepf1993} note a relative excess of irregular galaxy shapes in compact groups.  We discuss potential tidal effects on the lens in Section \ref{sec:fg_and_expot}. The maximum multipole coefficient (Section \ref{sec:mpoles}) is $b_3=0.0061$. This is within the regime of typical quadrupole strengths observed in numerical galaxy simulations by \cite{kochanek2004}, and boxy/disky features are not unusual in early-type galaxies.

\subsubsection{Stellar mass}
\label{sec:stellarmass}

We find consistent stellar masses for both composite models containing \sersic components; with $M_s\approx 1.5\times 10^{11}~M_\odot$. While this is a plausible stellar mass for a massive elliptical lens galaxy \citep[e.g.][]{auger2010}, dedicated follow-up observations in the optical/IR would be required for an independent constraint. The Keck AO observation used here has no absolute flux calibration, and a previous attempt at photometric modeling of the lens galaxy from HST observations was unsuccessful \citep{kochanek2000}.   

\subsubsection{Field galaxies and external potential}
\label{sec:fg_and_expot}

We also find that the PL+FG$_\mathrm{fixed}$ model, \revisions{with} masses of group galaxies fixed to the values derived by \cite{spingola2018}, \revisions{is relatively poor in} explaining the data.  Allowing the field galaxy masses to vary freely results in a drastically different gravitational environment. We show convergence maps for both PL+FG$_\mathrm{fixed}$ and PL+FG$_\mathrm{free}$ in a $15\times15$~\revisions{arcsec$^2$} region in \revisions{Fig.}~\ref{fig:model2}. Model PL+FG$_\mathrm{free}$ increases the mass of the nearby BCG (G1) by a factor of 3. The masses of G6 and G4 increase slightly, but G5 and G2 disappear altogether.  It seems that for the moment, the existing optical/IR observations of this group cannot alone inform a satisfactory mass model.  As a rule, coordinated multi-instrument observations are needed to adequately constrain the mass distribution in such complex environments \citep[e.g.][]{lagattuta2017,sluse2017,sluse2019,montes2019}. We therefore exclude the FG models from the rest of the discussion, as they contain too much uncertainty in the external convergence of the field galaxies. 

In light of this result, we instead considered the environment of this lens in terms of a third-order expansion of the external potential around G3 (Section \ref{sec:expot}). Model PL+EP gives an improvement in the Bayesian log-evidence of 536 relative to PL+FG$_\mathrm{free}$, with fewer free parameters; we therefore deem the EP model as having sufficient complexity to capture the effects of the local gravitational landscape, but in a more generic parameterization.  \cite{bernstein1999} note that for spherically-symmetric cluster potentials, $\tau \sim \delta \sim \Gamma^2$ and $\theta_\tau \sim \theta_\delta \sim \theta_\Gamma$.  We find that for all EP models, $\Gamma\approx0.08$, with $\tau$ and $\delta \lesssim 0.05$, $\theta_\Gamma\approx 75$ deg, and $\theta_\delta\approx 55$ deg. $\theta_\tau$ varies between $-14$ and $-65$ deg. As the mass distribution of this galaxy group is clearly quite far from spherical symmetry, our results for the external potential are still plausible. 

The EP model is intended to capture gravitational effects solely from external sources. However, given the wide variation in $\theta_\tau$, the extent to which EP may be degenerate with internal properties of the galaxy that multipole and/or \sersic components fail to capture is unclear, given the data available to us. We note the presence of mild correlations between, \revisions{for example}, $\theta_\delta$ and the internal multipole coefficients (see the posteriors in \revisions{Figs.}~\ref{fig:post_m4_ep} and \ref{fig:post_m4_sr_ep}), indicating that there is some interplay between nominally ``internal'' and ``external'' degrees of freedom in the lens model.

\subsection{Time-delay cosmography}

 Measurements of the Hubble constant ($H_0$) inferred with time-delay cosmography are known to be biased if the mass profile of the lens is not sufficiently well-known \citep{schneider2013,xu2016,xu2017,kochanek2020,enzi2020,birrer2021}.  In the absence of detailed, high-quality models for the external convergence \citep{rusu2017,sluse2017,birrer2019,tihh2020} or spatially-resolved kinematics \citep{birrer2021,yil2021}, we cannot make precise claims regarding an absolute measurement of $H_0$ using this lens. Instead, we will consider the fractional bias $f_{H_0}$ in a measurement of $H_0$ relative to our best model PL+MP+SR. We compute this bias using the relation given by \cite{kochanek2020}, which describes how the inferred $H_0$ relates to the convergence at the Einstein radius of the lens:
\begin{equation} \label{eq:tdc}
    f_{H_0} = \frac{H_{0,\mathrm{true}}}{H_{0,\mathrm{model}}}-1 = \frac{1-\kappa_{E,\mathrm{true}}}{1-\kappa_{E,\mathrm{model}}}-1,
\end{equation}
where $\kappa_E \equiv \kappa(R_E)$ is the convergence at the Einstein radius of the lens.  We compute $R_E$ and $\kappa_E$ numerically from the total convergence maps. The results are shown in Table \ref{tab:logev}. We find that the largest bias in $H_0$, at $f_{H_0}=10.3$ per cent, comes from model PL+SR+EP.  The inclusion of angular multipoles in model PL+MP+SR+EP substantially improves on PL+SR+EP, with $f_{H_0} = 4.8$ per cent.  This example highlights the importance of including sufficient angular complexity in the lens model when making this type of measurement.  The necessity for sufficient angular structure in lens models has also been identified by \citet{kochanek2021} and \citet{cao2022} for individual systems. \revisions{\citet{vdv2022} find that omission of multipole structure from lens models can bias measurements of $H_0$ for individual systems, similarly to what we observe for \jzsfo. However, when considering a population of lenses, they find that the inference on $H_0$ remains unbiased, albeit with extra uncertainty. Aside from azimuthal structures, our results show that lens galaxies can also have complex radial structures, which are expected to lead to systematic biases even on a population level.}

\subsection{Flux-ratio anomalies}

\begin{landscape}
 \begin{figure}
 \centering
   \includegraphics[scale=\figscale]{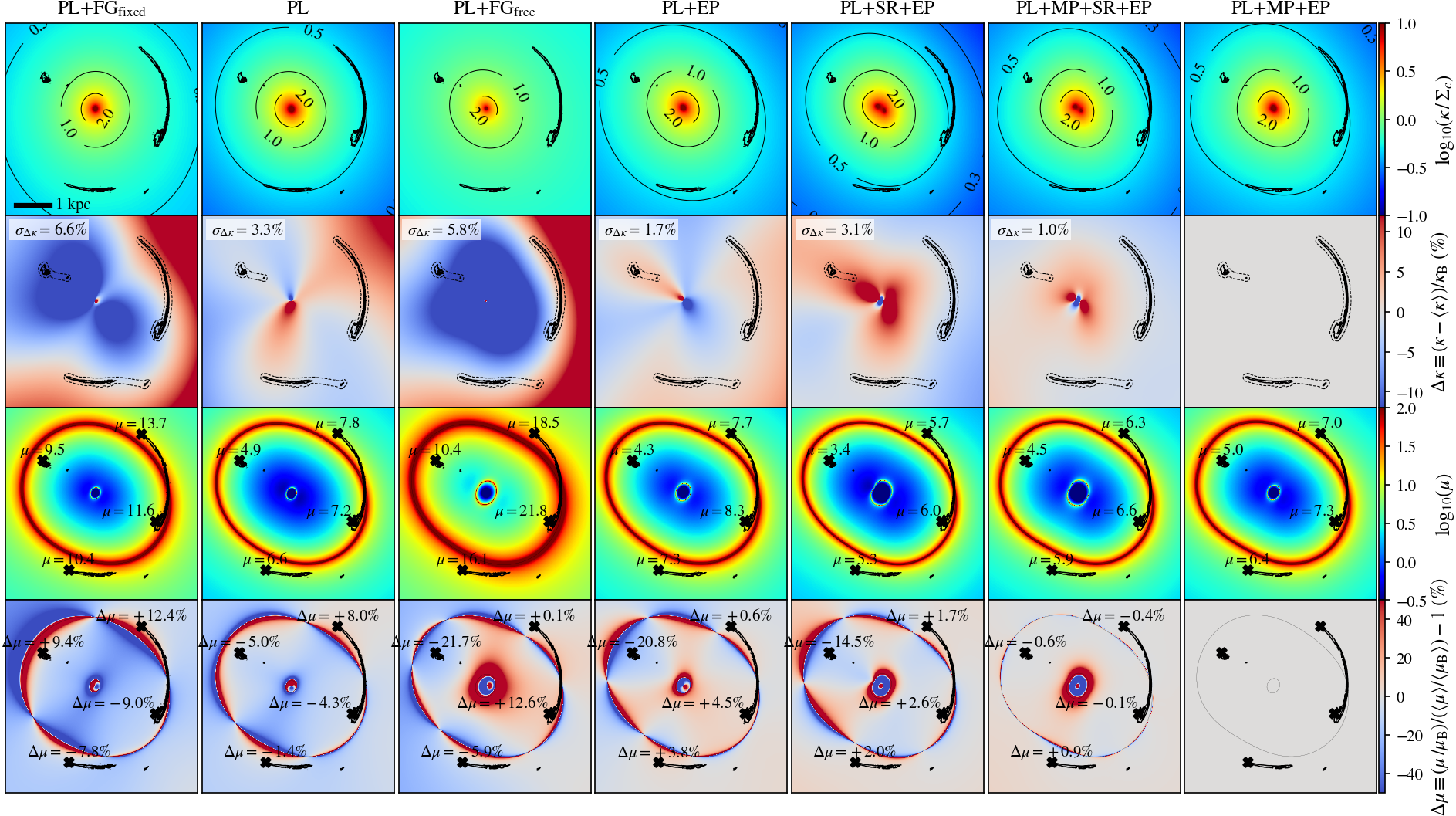}  \caption{ Comparison of convergence and magnification maps between lens models, ordered from left to right by increasing Bayesian evidence. The physical extent is the same for all panels; a scale bar is shown in the top-left panel. The colour scales are consistent within each row. Top row: Total convergence maps, in units of critical density $\Sigma_c$. Second row: RMS differences from the best model PL+MP+SR as a fraction of the total convergence, computed inside the light mask (Section \ref{sec:mask}). Third row: Magnification maps, with magnifications of the bright source component 1 (see \revisions{Fig.}~\ref{fig:modelcmp}) labeled for each of the four images. Bottom row: Differences in magnification relative to PL+MP+SR. We additionally plot the change in magnification at the location of the brightest image corresponding to source component 1. Relative magnifications are standardized to the flux-weighted mean magnification for each model, in order to remove the mass-sheet degeneracy from the comparison. }
  \label{fig:magcmp}
\end{figure}
\end{landscape}

We also assess the impact of the assumed lens profile on the measured flux-ratios of the lensed images.  We remove the mass sheet degeneracy by first normalizing the magnifications to the flux-weighted mean for each model. We then compute the observed magnifications at the image positions corresponding to the brightest source component (component 1; see \revisions{Fig.}~\ref{fig:modelcmp} and \ref{fig:magcmp}\revisions{)} for each of the lens models. We find that the maximum change in magnification for any of these images (relative to PL+MP+EP) is 20.8 per cent for the PL+EP model, and on the order of 5 \revisions{to} 15 per cent in general (again excluding the FG models). Such large changes are comparable to the scatter in the measured flux ratios from gravitational lens systems, which are typically attributed to unconstrained mass structure in the lens \citep{xu2015,hsueh2016,hsueh2017,hsueh2018b}.  This demonstrates that density structure in the lens beyond a PEMD is a plausible source of this scatter, and that the density profile must depart from a PEMD by $\lesssim3$ per cent locally in order to produce non-negligible flux-ratio anomalies.  Our results support the conclusions of \cite{gilman2017}, \cite{hsueh2018b} and \cite{he2022}, who find that \revisions{a good} understanding of the galaxy-scale mass structure of a lens is paramount for robustly inferring properties of the dark matter subhalo population, using either lensed quasars or galaxies.

\subsection{Angular resolution}

To illustrate the importance of high resolution imaging in differentiating between lens models, we again compare the lens models using the Keck AO observation of \jzsfo (Section \ref{sec:aodata}). When both the lens and the source are left free to vary, we find that models more complex than the PL cannot be constrained at all. We therefore keep the parameters of each model fixed at the best values inferred from the VLBI data and re-optimize only for the lens position and the source regularization strength, comparing them in terms of their log-evidence.

We find that their order differs from the model ranking using the VLBI data, but that the log-evidence values are much closer. Setting $\Delta \logev_\mathrm{PL+MP+EP}\equiv 0$, we find $\Delta \logev_\mathrm{PL+MP+SR+EP} = +17$, $\Delta \logev_\mathrm{PL+SR+EP} = +17$, $\Delta \logev_\mathrm{PL+EP} = +5$, $\Delta \logev_\mathrm{PL} = 0$, $\Delta \logev_\mathrm{PL+FG_{free}} = +19$, and $\Delta \logev_\mathrm{PL+FG_{fixed}} = -46$. These model differences are at least two orders of magnitude weaker than those obtained from the VLBI data. 
The maximum difference between either of these two models using the Keck AO data is $\Delta \logev_\mathrm{max,AO}=65$, while $\Delta \logev_\mathrm{max,VLBI}=22043$ for the VLBI data.  Therefore, the ability to resolve source structure in the lensed images on \revisions{mas-scales, with either VLBI at radio to mm wavelengths or with Extremely Large Telescopes (ELTs) in the future,} is of major consequence for sensitivity to the mass structure in lens galaxies.

\section{Conclusions} \label{sec:conclusion}

In this paper, we have demonstrated the power of high-resolution VLBI observations in constraining the mass density profile of the gravitational lens system \jzsfo, and in particular the ability to differentiate a preference in the data for different types of mass structure. 

As demonstrated in \revisions{Fig.}~\ref{fig:zoomcmp}, misalignments between source images on the order of milliarcseconds are made obvious. These differences, which are not detectable with lower-resolution data, are key to revealing radial and angular complexity in the lens. In particular, we found that the simple and standard choice of an elliptical power-law mass density profile is a good fit to the data down to scales of $\sim$5 mas. This result has potentially important implications for the quick  modelling of the large samples of (relatively low resolution) data that will be provided in the future by surveys with {\it Euclid} and the Vera C. Rubin Observatory. 

Observed at scales smaller than $\sim$5 mas, the lens galaxy in \jzsfo shows significant structure that is best captured by a model including angular multipoles, a surface mass gradient, and a shear gradient, in addition to the power-law component. We have shown that ignoring the presence of these extra components has important implications for time-delay cosmography and the physical interpretation of flux-ratio anomalies. This is the first pixellated source model reconstructed from a gravitational lens system observed at such high angular resolution and SNR. Larger samples of gravitational lens systems with this data-quality will be required to understand the complexity of galaxies in the general lens population.

Our best-fitting model still fails to completely focus the source light, with misalignments between source images of the order of $\sim 1-2$ mas still present. However, we showed that a more complex parametric model, which also includes a \sersic profile for the  baryonic component of the lens galaxy, is significantly dis-preferred by the data. This result possibly indicates that there is extra complexity in the lens mass distribution that cannot be accounted for with relatively simple parametric prescriptions. A major next step forward for this research will be the extension of this analysis to include pixellated potential corrections \citep{koopmans2005,vegetti2008,suyu2009} in the lens model. We will present this analysis in a forthcoming paper.

Finally, we have also shown the computational feasibility of modeling large, high-resolution interferometric data sets using this method. The numerical techniques \revisions{derived by} \cite{powell2021}, along with the fast $\chi^2$ computation presented in Section \ref{sec:fastchi2}, drastically \revisions{speed up} the evaluation of the posterior samples. Such capabilities will be crucial \revisions{in the near term with LOFAR and in the future} during the era of the Square Kilometre Array (SKA), in which \revisions{$>10^5$} new radio lenses will be discovered \citep{koopmans2004,mckean2015}.  Follow-up of these lenses with VLBI and optical/IR instruments will provide unprecedented constraining power on the physical processes \revisions{that} set the density profiles in lens galaxies.

\begin{landscape}
\begin{table} 
\caption{ Values of inferred lens parameters $\etalens$ and source regularization strength $\lams$ for all lens models considered in this work (see Section \ref{sec:lensmodel}; Table \ref{tab:lensmodels}). We quote the mean and 95 per cent confidence interval, as well as the MAP parameter values.  Quoted confidence intervals for the lens parameters are estimated as described in Section \ref{sec:conv_mag_results}. }
\centering
\begin{tabular}{l l@{\hspace{0.2cm}}ll l@{\hspace{0.2cm}}ll l@{\hspace{0.2cm}}ll}
\hline
\noalign{\vskip 0.1cm}
 & \multicolumn{3}{c}{PL+FG$_\mathrm{fixed}$} & \multicolumn{3}{c}{PL} & \multicolumn{3}{c}{PL+FG$_\mathrm{free}$} \\
 \cmidrule(lr){2-4} \cmidrule(lr){5-7} \cmidrule(lr){8-10}
Par. & \multicolumn{2}{l}{Mean $\pm\,2\sigma$ (95\% CI)} & MAP & \multicolumn{2}{l}{Mean $\pm\,2\sigma$ (95\% CI)} & MAP & \multicolumn{2}{l}{Mean $\pm\,2\sigma$ (95\% CI)} & MAP \\
\cmidrule(lr){1-1} \cmidrule(lr){2-4} \cmidrule(lr){5-7} \cmidrule(lr){8-10}
\noalign{\vskip 0.05cm}
$\kappa_0$ & $0.371$&$\pm\,0.011$ & $0.371$ & $0.4627$&$\pm\,0.0028$ & $0.4624$ & $0.40254$&$\pm\,0.00054$ & $0.40241$\\
\noalign{\vskip 0.05cm}
$\theta_q~(^\circ)$ & $-1.2$&$\pm\,15.8$ & $-1.2$ & $19.17$&$\pm\,0.65$ & $19.16$ & $28.67$&$\pm\,0.31$ & $28.69$\\
\noalign{\vskip 0.05cm}
$q$ & $0.910$&$\pm\,0.054$ & $0.910$ & $0.8993$&$\pm\,0.0076$ & $0.8992$ & $0.88862$&$\pm\,0.00083$ & $0.88861$\\
\noalign{\vskip 0.05cm}
$x_0~(")$ & $-0.43739$&$\pm\,0.00086$ & $-0.43739$ & $-0.44153$&$\pm\,0.00041$ & $-0.44152$ & $-0.45271$&$\pm\,0.00006$ & $-0.45274$\\
\noalign{\vskip 0.05cm}
$y_0~(")$ & $0.1780$&$\pm\,0.0023$ & $0.1780$ & $0.17507$&$\pm\,0.00002$ & $0.17507$ & $0.17968$&$\pm\,0.00019$ & $0.17970$\\
\noalign{\vskip 0.05cm}
$\gamma$ & $1.903$&$\pm\,0.028$ & $1.903$ & $1.8977$&$\pm\,0.0047$ & $1.8982$ & $1.75854$&$\pm\,0.00086$ & $1.75821$\\
\noalign{\vskip 0.05cm}
$\Gamma$ & $0.0304$&$\pm\,0.0016$ & $0.0304$ & $0.0925$&$\pm\,0.0013$ & $0.0925$ & $0.07890$&$\pm\,0.00091$ & $0.07922$\\
\noalign{\vskip 0.05cm}
$\theta_\mathrm{\Gamma}~(^\circ)$ & $57.3$&$\pm\,25.6$ & $57.3$ & $73.70$&$\pm\,0.77$ & $73.70$ & $172.93$&$\pm\,0.21$ & $172.85$\\
\noalign{\vskip 0.05cm}
$\kappa_\mathrm{G1}$ &  $\equiv 0.751$ & &  &  - & &  & $2.3752$&$\pm\,0.0084$ & $2.3784$\\
\noalign{\vskip 0.05cm}
$\kappa_\mathrm{G2}$ &  $\equiv 0.266$ & &  &  - & &  & $0.00025$&$\pm\,0.00047$ & $0.00004$\\
\noalign{\vskip 0.05cm}
$\kappa_\mathrm{G4}$ &  $\equiv 0.341$ & &  &  - & &  & $0.554$&$\pm\,0.021$ & $0.558$\\
\noalign{\vskip 0.05cm}
$\kappa_\mathrm{G5}$ &  $\equiv 0.435$ & &  &  - & &  & $0.00029$&$\pm\,0.00053$ & $0.00016$\\
\noalign{\vskip 0.05cm}
$\kappa_\mathrm{G6}$ &  $\equiv 0.282$ & &  &  - & &  & $0.4905$&$\pm\,0.0054$ & $0.4912$\\
\noalign{\vskip 0.05cm}
$\lambda_s~(\times 10^9)$ & $2.834$&$\pm\,0.044$ & $2.830$ & $4.247$&$\pm\,0.070$ & $4.261$ & $5.74$&$\pm\,0.09$ & $5.77$\\
\noalign{\vskip 0.05cm}
\noalign{\vskip 0.05cm}
\hline
\noalign{\vskip 0.5cm}
\end{tabular}
\\
\begin{tabular}{l l@{\hspace{0.2cm}}ll l@{\hspace{0.2cm}}ll l@{\hspace{0.2cm}}ll l@{\hspace{0.2cm}}ll}
\hline
\noalign{\vskip 0.1cm}
 & \multicolumn{3}{c}{PL+EP} & \multicolumn{3}{c}{PL+SR+EP} & \multicolumn{3}{c}{PL+MP+SR+EP} & \multicolumn{3}{c}{PL+MP+EP} \\
 \cmidrule(lr){2-4} \cmidrule(lr){5-7} \cmidrule(lr){8-10} \cmidrule(lr){11-13}
Par. & \multicolumn{2}{l}{Mean $\pm\,2\sigma$ (95\% CI)} & MAP & \multicolumn{2}{l}{Mean $\pm\,2\sigma$ (95\% CI)} & MAP & \multicolumn{2}{l}{Mean $\pm\,2\sigma$ (95\% CI)} & MAP & \multicolumn{2}{l}{Mean $\pm\,2\sigma$ (95\% CI)} & MAP \\
\cmidrule(lr){1-1} \cmidrule(lr){2-4} \cmidrule(lr){5-7} \cmidrule(lr){8-10} \cmidrule(lr){11-13}
\noalign{\vskip 0.05cm}
$\kappa_0$ & $0.4976$&$\pm\,0.0018$ & $0.4976$ & $0.2934$&$\pm\,0.0027$ & $0.2935$ & $0.3527$&$\pm\,0.0033$ & $0.3522$ & $0.4793$&$\pm\,0.0026$ & $0.4792$\\
\noalign{\vskip 0.05cm}
$\theta_q~(^\circ)$ & $33.56$&$\pm\,0.11$ & $33.61$ & $63.16$&$\pm\,0.37$ & $62.96$ & $63.65$&$\pm\,1.18$ & $63.06$ & $28.51$&$\pm\,0.27$ & $28.51$\\
\noalign{\vskip 0.05cm}
$q$ & $0.87468$&$\pm\,0.00059$ & $0.87498$ & $0.8741$&$\pm\,0.0019$ & $0.8756$ & $0.92391$&$\pm\,0.00091$ & $0.92397$ & $0.8741$&$\pm\,0.0026$ & $0.8738$\\
\noalign{\vskip 0.05cm}
$x_0~(")$ & $-0.44627$&$\pm\,0.00009$ & $-0.44626$ & $-0.45903$&$\pm\,0.00034$ & $-0.45883$ & $-0.45564$&$\pm\,0.00051$ & $-0.45544$ & $-0.44217$&$\pm\,0.00029$ & $-0.44222$\\
\noalign{\vskip 0.05cm}
$y_0~(")$ & $0.18364$&$\pm\,0.00010$ & $0.18368$ & $0.19131$&$\pm\,0.00037$ & $0.19099$ & $0.18914$&$\pm\,0.00039$ & $0.18892$ & $0.18019$&$\pm\,0.00032$ & $0.18014$\\
\noalign{\vskip 0.05cm}
$\gamma$ & $1.8410$&$\pm\,0.0029$ & $1.8409$ & $1.8754$&$\pm\,0.0046$ & $1.8775$ & $1.8172$&$\pm\,0.0048$ & $1.8206$ & $1.8707$&$\pm\,0.0043$ & $1.8709$\\
\noalign{\vskip 0.05cm}
$a_3$ &  - & &  &  - & &  & $-0.00122$&$\pm\,0.00009$ & $-0.00123$ & $-0.00219$&$\pm\,0.00012$ & $-0.00218$\\
\noalign{\vskip 0.05cm}
$b_3$ &  - & &  &  - & &  & $-0.00608$&$\pm\,0.00006$ & $-0.00607$ & $-0.00501$&$\pm\,0.00016$ & $-0.00498$\\
\noalign{\vskip 0.05cm}
$a_4$ &  - & &  &  - & &  & $0.00023$&$\pm\,0.00003$ & $0.00024$ & $0.00150$&$\pm\,0.00007$ & $0.00151$\\
\noalign{\vskip 0.05cm}
$b_4$ &  - & &  &  - & &  & $0.00088$&$\pm\,0.00005$ & $0.00088$ & $0.00159$&$\pm\,0.00006$ & $0.00160$\\
\noalign{\vskip 0.05cm}
$\Gamma$ & $0.07808$&$\pm\,0.00032$ & $0.07800$ & $0.08734$&$\pm\,0.00045$ & $0.08761$ & $0.08524$&$\pm\,0.00038$ & $0.08549$ & $0.08698$&$\pm\,0.00070$ & $0.08703$\\
\noalign{\vskip 0.05cm}
$\theta_\mathrm{\Gamma}~(^\circ)$ & $77.044$&$\pm\,0.090$ & $76.991$ & $74.388$&$\pm\,0.074$ & $74.351$ & $72.320$&$\pm\,0.092$ & $72.373$ & $77.08$&$\pm\,0.28$ & $77.11$\\
\noalign{\vskip 0.05cm}
$\tau$ & $0.0486$&$\pm\,0.0011$ & $0.0492$ & $0.0594$&$\pm\,0.0011$ & $0.0592$ & $0.0573$&$\pm\,0.0011$ & $0.0570$ & $0.0457$&$\pm\,0.0041$ & $0.0450$\\
\noalign{\vskip 0.05cm}
$\theta_\tau~(^\circ)$ & $-14.23$&$\pm\,1.10$ & $-14.37$ & $-64.37$&$\pm\,1.47$ & $-65.20$ & $-45.68$&$\pm\,1.09$ & $-46.43$ & $-54.03$&$\pm\,2.68$ & $-54.25$\\
\noalign{\vskip 0.05cm}
$\delta$ & $0.02536$&$\pm\,0.00015$ & $0.02541$ & $0.02097$&$\pm\,0.00025$ & $0.02111$ & $0.04973$&$\pm\,0.00036$ & $0.04971$ & $0.04415$&$\pm\,0.00096$ & $0.04397$\\
\noalign{\vskip 0.05cm}
$\theta_\delta~(^\circ)$ & $56.84$&$\pm\,0.13$ & $56.77$ & $50.01$&$\pm\,0.18$ & $49.94$ & $52.15$&$\pm\,0.26$ & $52.13$ & $51.56$&$\pm\,0.48$ & $51.58$\\
\noalign{\vskip 0.05cm}
$M_s (10^{10}~M_\odot)$ &  - & &  & $15.52$&$\pm\,0.17$ & $15.44$ & $12.68$&$\pm\,0.29$ & $12.60$ &  - & & \\
\noalign{\vskip 0.05cm}
$\lambda_s~(\times 10^9)$ & $6.03$&$\pm\,0.10$ & $6.01$ & $7.33$&$\pm\,0.13$ & $7.30$ & $10.54$&$\pm\,0.13$ & $10.52$ & $10.74$&$\pm\,0.20$ & $10.83$\\
\noalign{\vskip 0.05cm}
\noalign{\vskip 0.05cm}
\hline
\end{tabular}
\label{tab:lenspars}
\end{table}
\end{landscape}

\section*{Acknowledgements}

\revisions{We thank Lyne Van de Vyvere and the referee, Simon Dye, for their insightful feedback and discussions.} DP, SV, and HRS acknowledge funding from the European Research Council (ERC) under the European Union's Horizon 2020 research and innovation programme (LEDA: grant agreement No 758853). SV thanks the Max Planck Society for support through a Max Planck Lise Meitner Group. JPM acknowledges support from the Netherlands Organization for Scientific Research (NWO) (Project No. 629.001.023) and the Chinese Academy of Sciences (CAS) (Project No. 114A11KYSB20170054). 
CS acknowledges financial support from the Italian Ministry of University and Research - Project Proposal CIR01\_00010. CDF acknowledges support for this work from the National Science Foundation under Grant No. AST-1715611.

This research used SciPy, NumPy and Matplotlib packages for Python \citep{Virtanen:2020,Harris:2020,Hunter:2007}.   \revisions{Corner plots were generated using \software{GetDist} \citep{lewis2019}.}

\subsection*{Data Availability}

This paper makes use of the following EVN, VLBA and GBT data: GM070, available at http://archive.jive.nl/scripts/portal.php.
The Keck data are available from the Keck Observatory Archive at \texttt{https://www2.keck.hawaii.edu/koa/public/koa.php}.

\subsection*{Code Availability}

The lens modelling code used for the analysis is fully explained in \citet{powell2021}; see also \citet{vegetti2008}, \citet{rybak2015b}, \citet{rizzo2018} and \citet{ritondale2019} for further details. The reader interested in using this code should contact the corresponding author.

%%%%%%%%%%%%%%%%%%%%%%%%%%%%%%%%%%%%%%%%%%%%%%%%%%

\bibliographystyle{mnras}
\bibliography{references}

%%%%%%%%%%%%%%%%%%%%%%%%%%%%%%%%%%%%%%%%%%%%%%%%%%

%%%%%%%%%%%%%%%%% APPENDICES %%%%%%%%%%%%%%%%%%%%%

\appendix

\section{Lens parameters} \label{app:lenspars}

\begin{figure*}
 \centering
   \includegraphics[scale=\cornerscale]{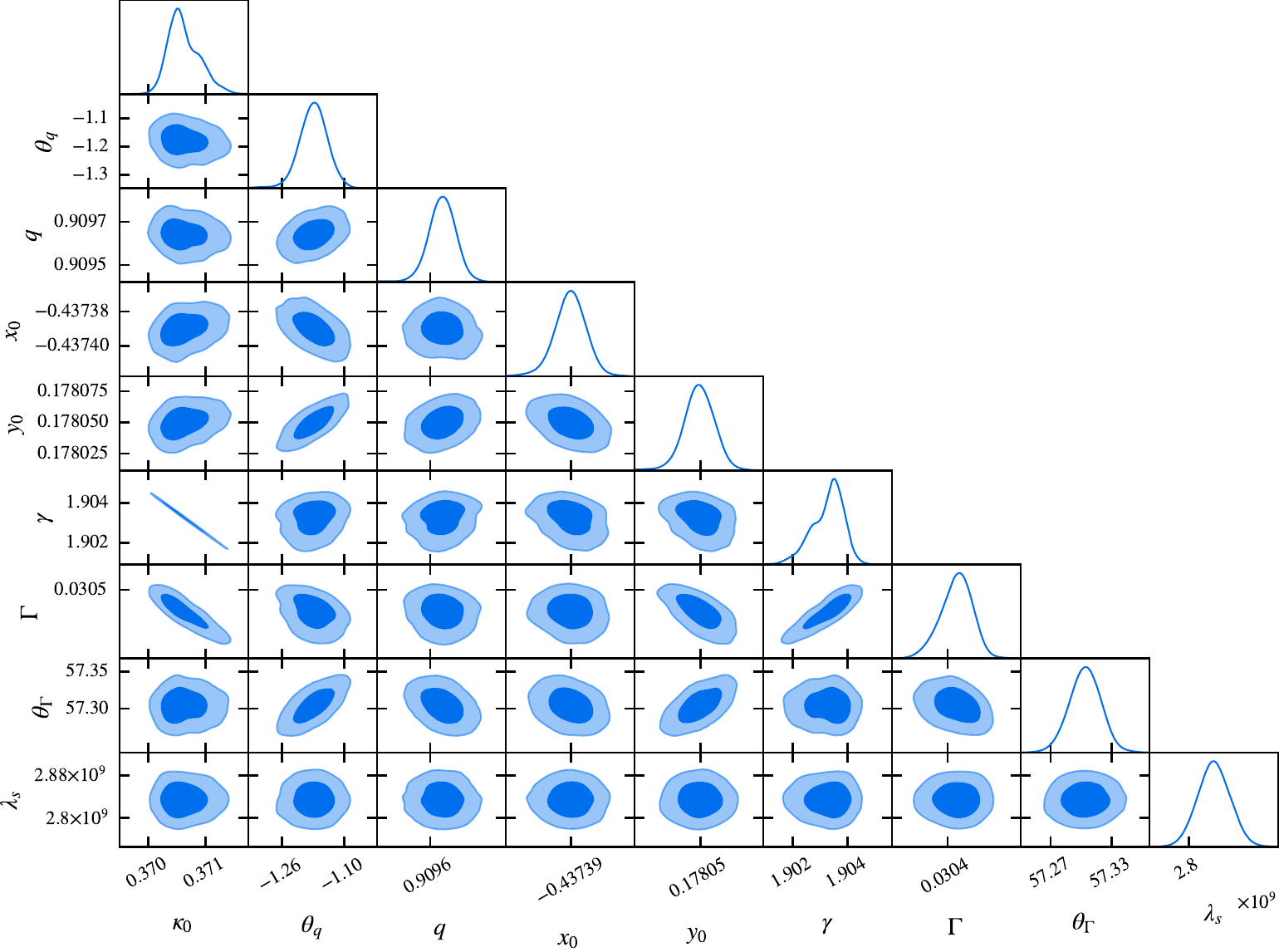} 
   \caption{Posterior distribution of parameters for model PL+FG$_\mathrm{fixed}$.} 
  \label{fig:post_fg_fixed}
\end{figure*}

\begin{figure*}
 \centering
   \includegraphics[scale=\cornerscale]{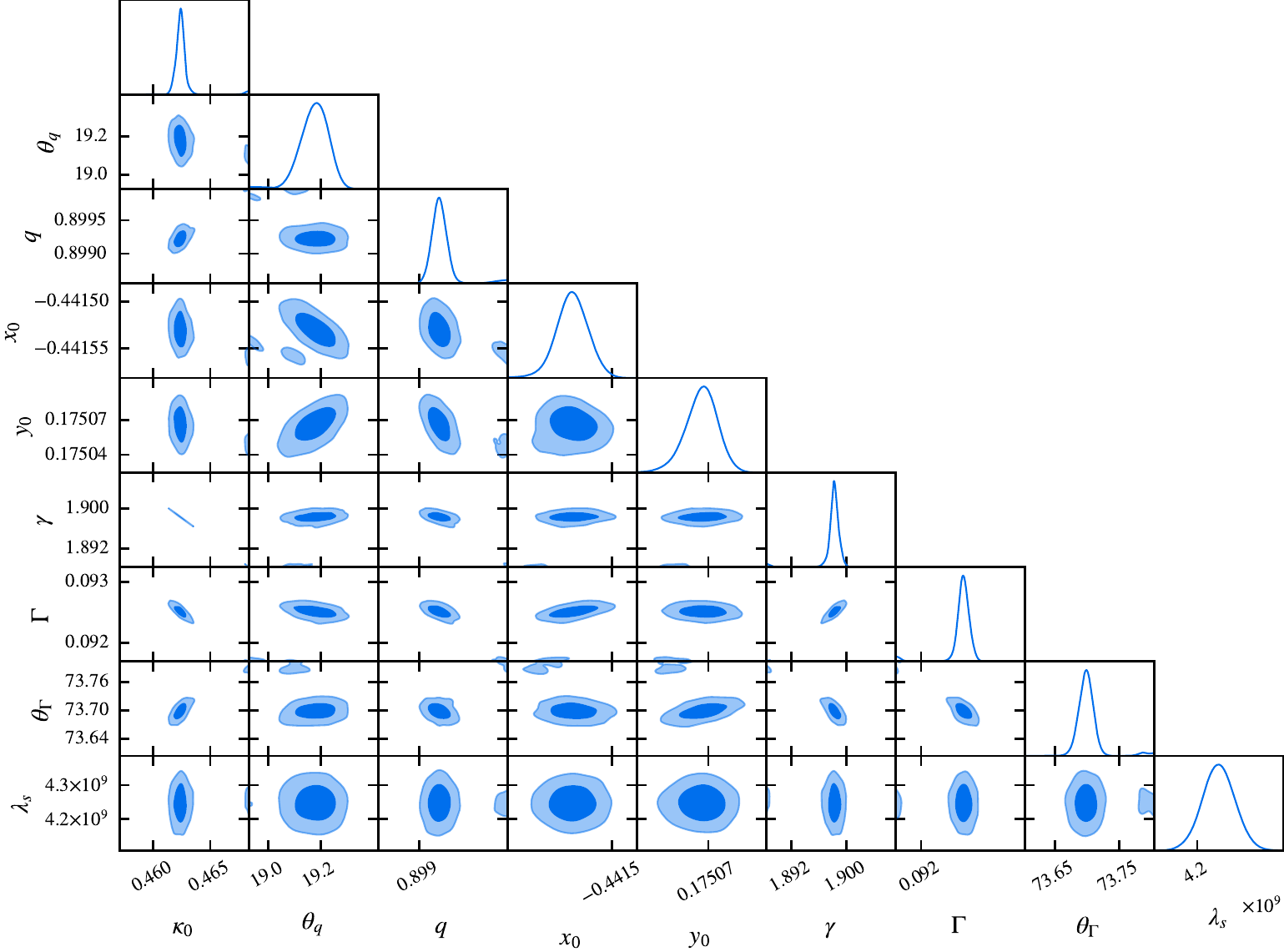}  
   \caption{Posterior distribution of parameters model PL. }
  \label{fig:post_pl}
\end{figure*}

\begin{landscape}
 \begin{figure}
 \centering
   \includegraphics[scale=\cornerscale]{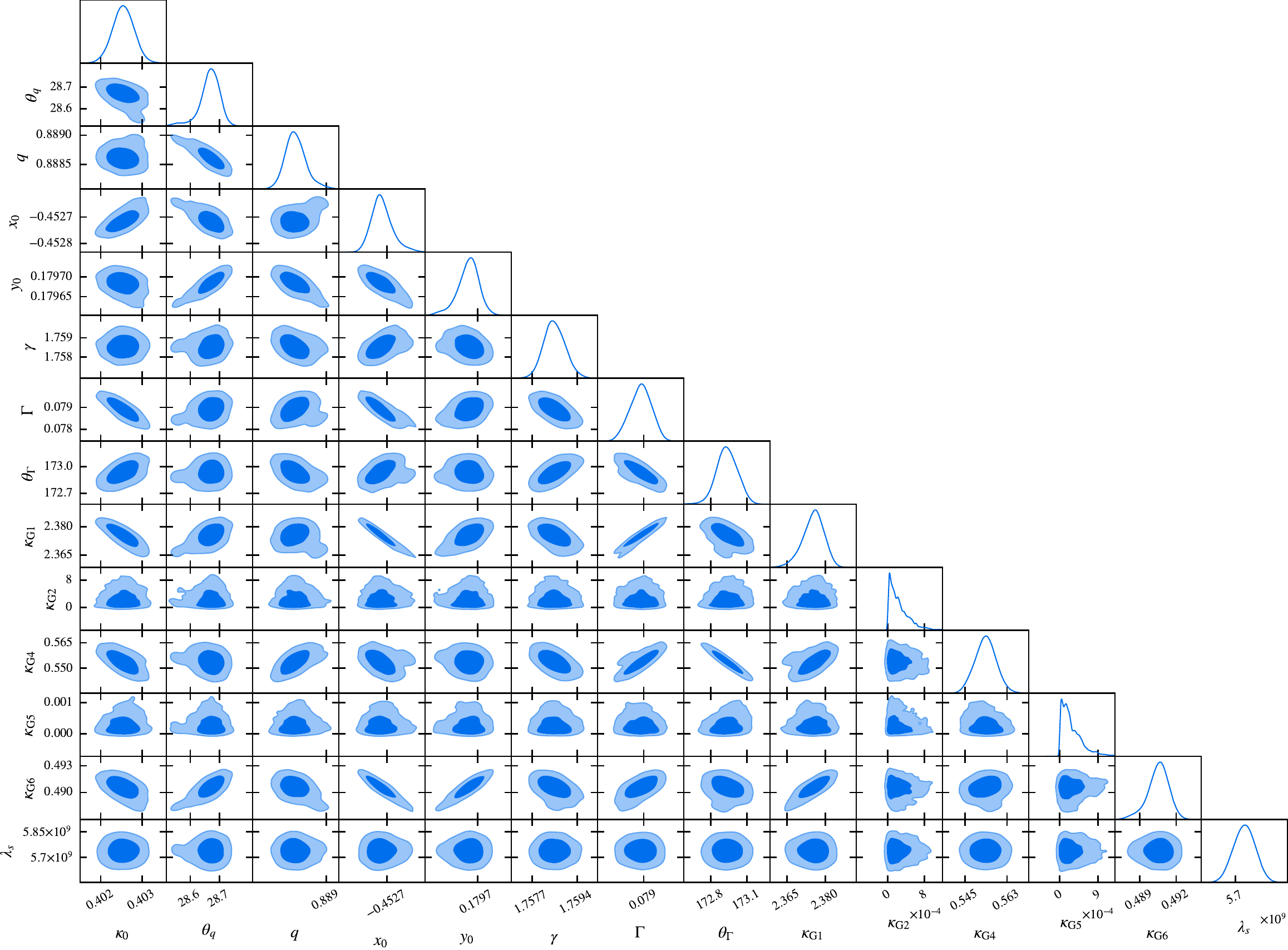} 
   \caption{Posterior distribution of parameters for model PL+FG$_\mathrm{free}$.} 
  \label{fig:post_fg_free}
\end{figure}
\end{landscape}

\begin{landscape}
 \begin{figure}
 \centering
   \includegraphics[scale=\cornerscale]{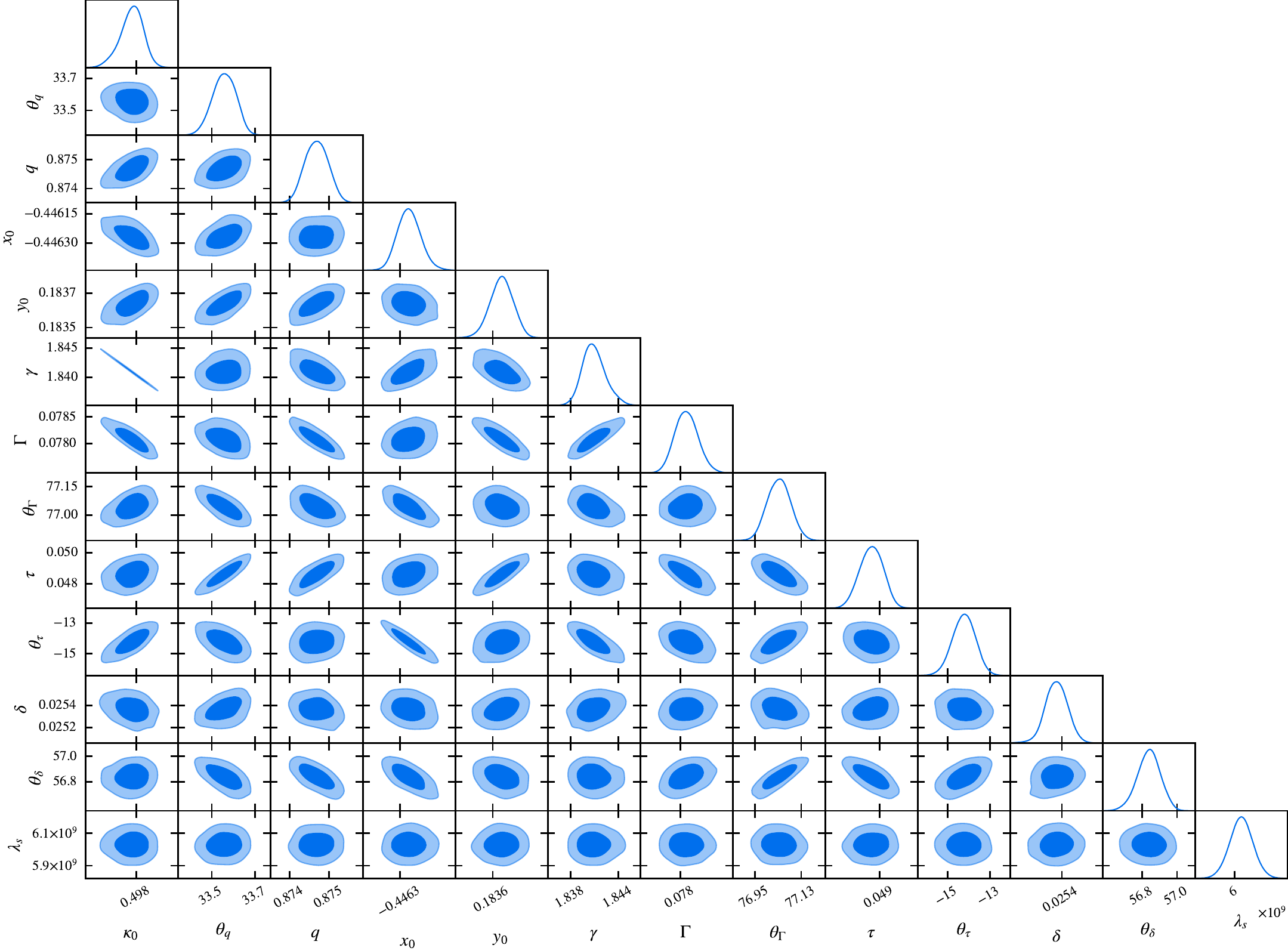}  
   \caption{Posterior distribution of parameters for model PL+EP. }
  \label{fig:post_pl_ep}
\end{figure}
\end{landscape}

\begin{landscape}
 \begin{figure}
 \centering
   \includegraphics[scale=\cornerscale]{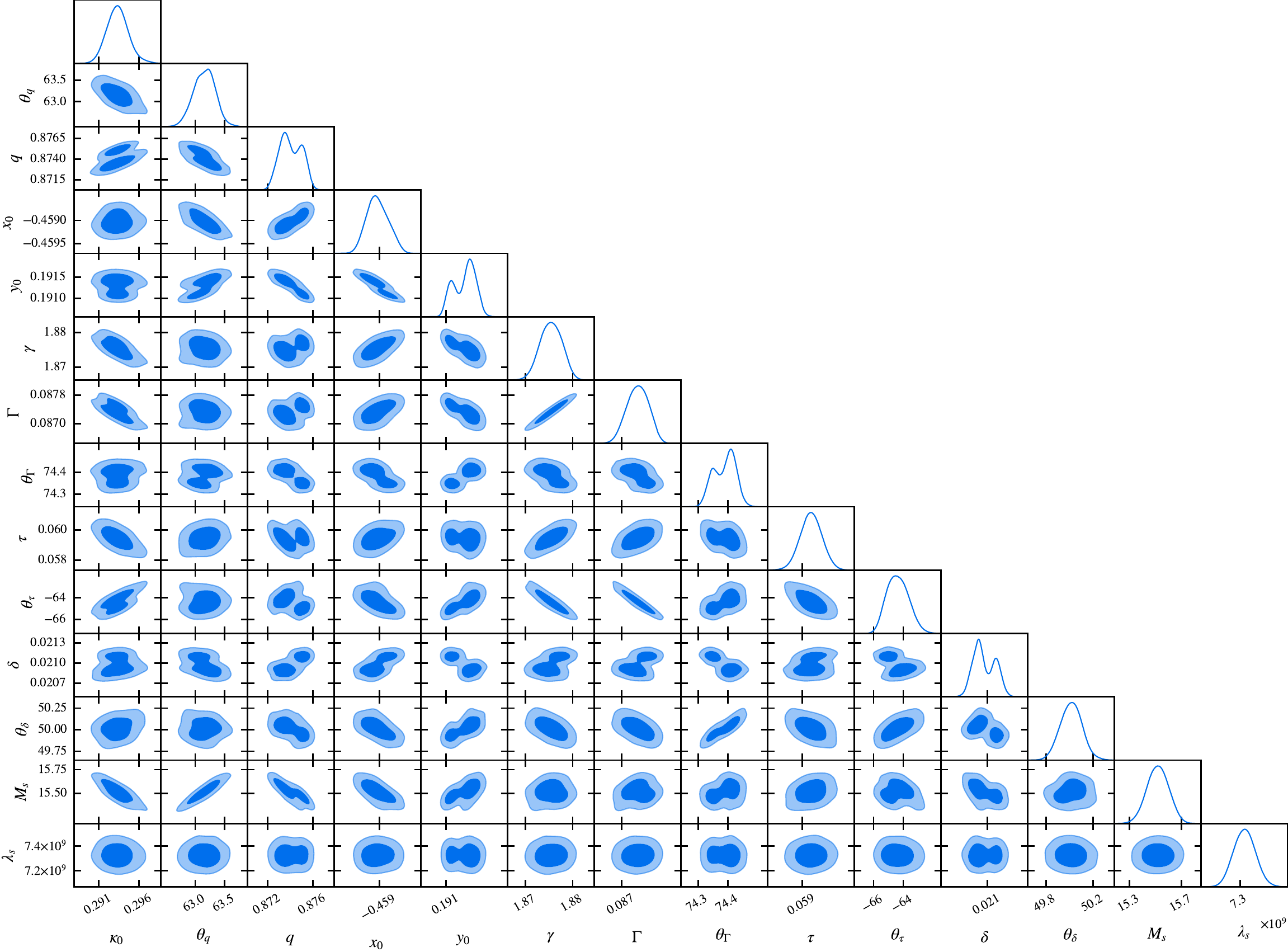}  
   \caption{Posterior distribution of parameters for model PL+SR+EP. }
  \label{fig:post_pl_sr_ep}
\end{figure}
\end{landscape}

\begin{landscape}
 \begin{figure}
 \centering
   \includegraphics[scale=\cornerscale]{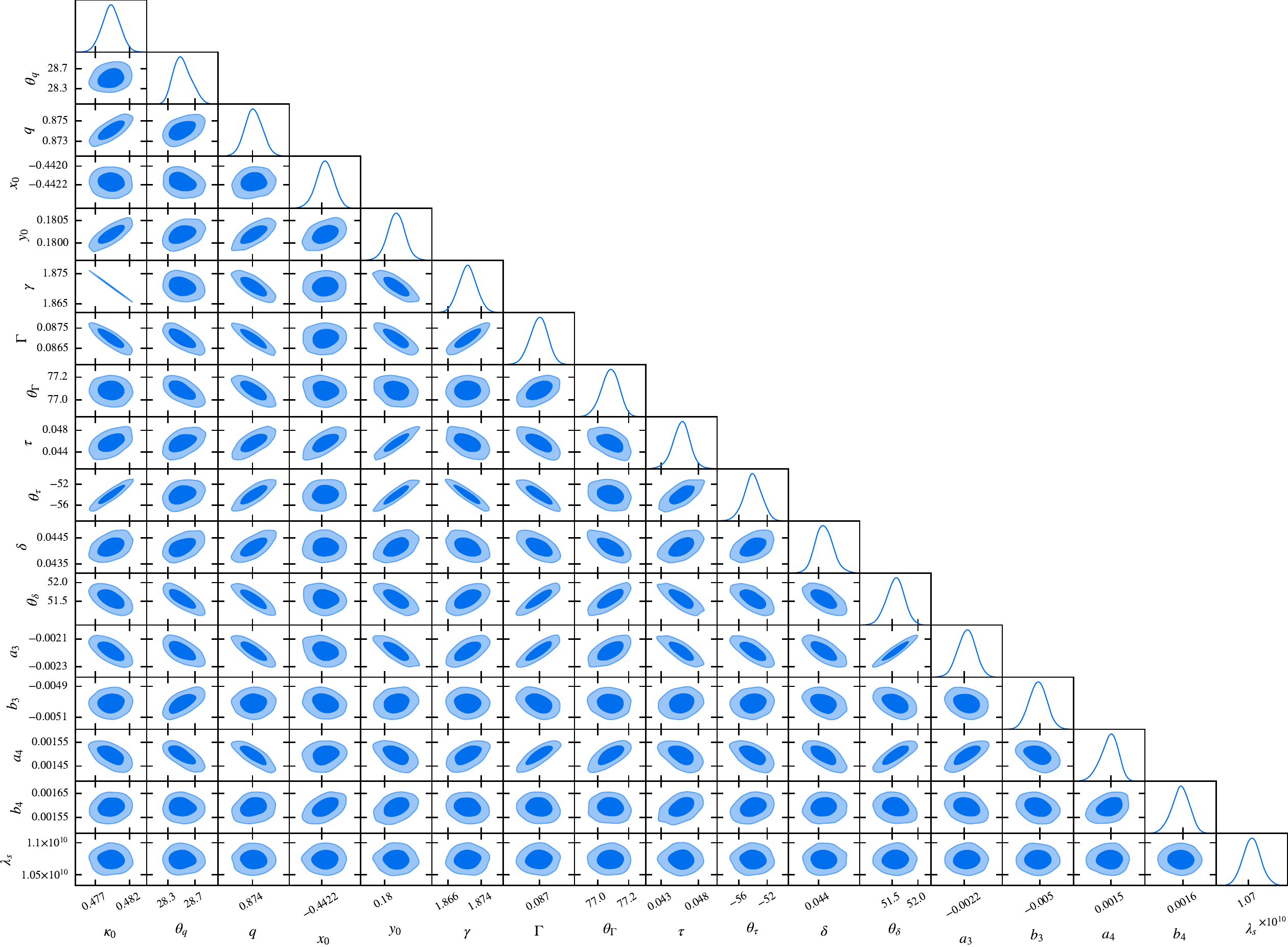}  
   \caption{Posterior distribution of parameters for model PL+MP+EP. }
  \label{fig:post_m4_ep}
\end{figure}
\end{landscape}

\begin{landscape}
 \begin{figure}
 \centering
   \includegraphics[scale=\cornerscale]{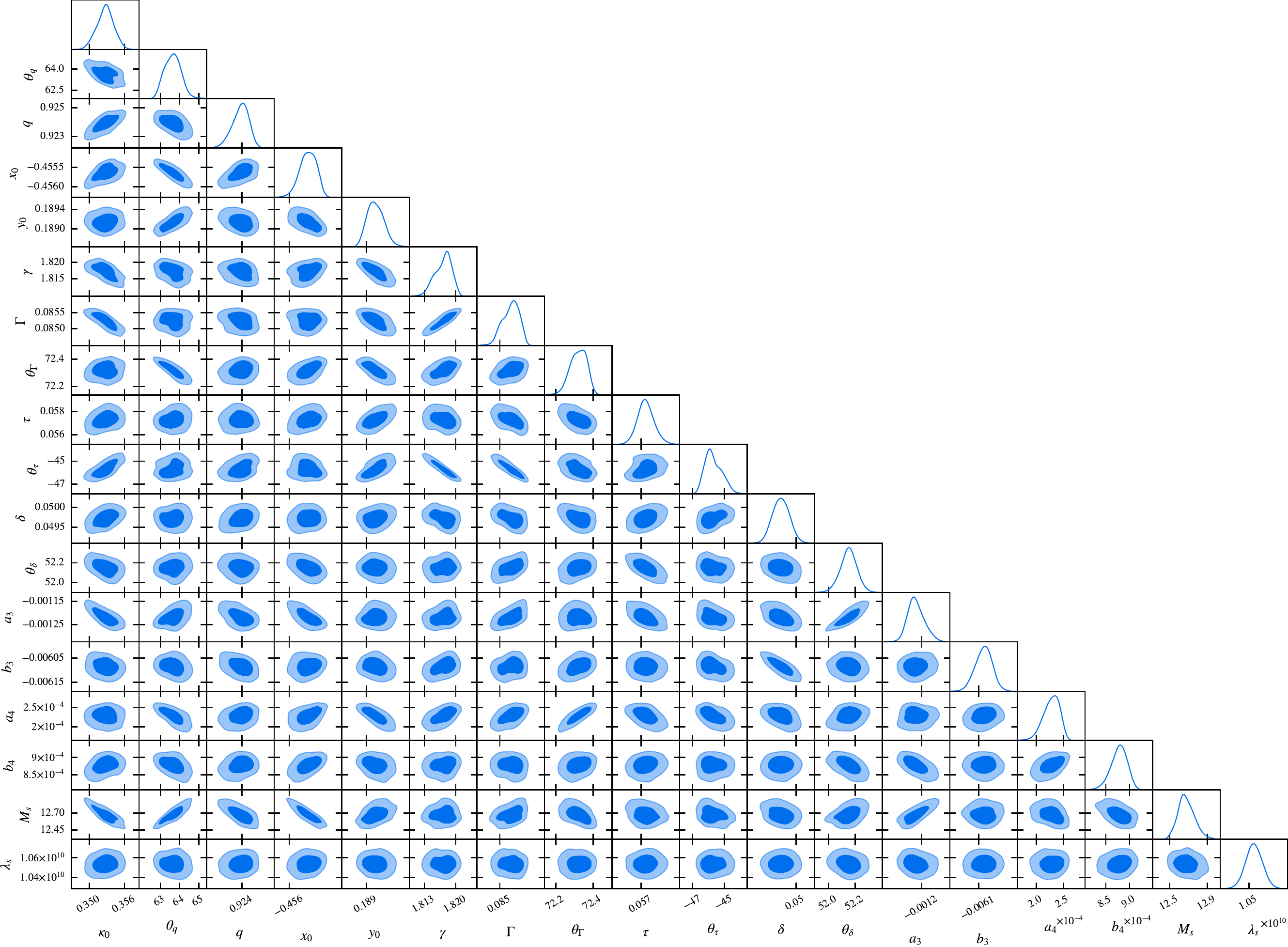}  
   \caption{Posterior distribution of parameters for model PL+MP+SR+EP. }
  \label{fig:post_m4_sr_ep}
\end{figure}
\end{landscape}

In this appendix, we present the full posterior distributions for the lens mass models studied in this paper. 
Table \ref{tab:lenspars} summarizes the parameters for each model considered in this work, along with the 95 per cent confidence intervals. Corner plots of the posteriors are shown in Figures \ref{fig:post_fg_fixed}, \ref{fig:post_pl}, \ref{fig:post_fg_free}, \ref{fig:post_pl_ep}, \ref{fig:post_pl_sr_ep}, \ref{fig:post_m4_ep}, and \ref{fig:post_m4_sr_ep}.

%%%%%%%%%%%%%%%%%%%%%%%%%%%%%%%%%%%%%%%%%%%%%%%%%%

% Don't change these lines
\bsp	% typesetting comment
\label{lastpage}
\end{document}